\documentclass[11pt]{article}
\pdfoutput=1

\usepackage{jheppub}
\usepackage{graphicx,color,verbatim,acronym,epsfig,subfigure,slashed,skak}
\usepackage{latexsym,amsfonts,amssymb,amsmath,makeidx,mathrsfs,multirow,lineno,bm,bbm,fancyvrb,ifpdf}

\newcommand{\GeV}      {~\mathrm{GeV}}
\newcommand{\TeV}      {~\mathrm{TeV}}

\newcommand{\pb}      {~\mathrm{pb}}
\newcommand{\fb}      {~\mathrm{fb}}
\newcommand{\ab}      {~\mathrm{ab}}

\def\beqn{\begin{eqnarray}}
\def\eeqn{\end{eqnarray}}
\def\beqs{\begin{subequations}}
\def\eeqs{\end{subequations}}
\def\beq{\begin{equation}}
\def\eeq{\end{equation}}
\def\ba{\begin{array}}
\def\ea{\end{array}}

\def\non{\nonumber\\}
\def\hf{\frac{1}{2}}

\def\mL{\mathcal{L}}
\def\mM{\mathcal{M}}

\def\mO{\mathcal{O}}

\def\mR{\mathcal{R}}

\title{Higgs pair productions in the CP-violating two-Higgs-doublet model}

\author[\symrook]{~Ligong Bian}
\emailAdd{lgbycl@cqu.edu.cn}
\affiliation[\symrook]{ Department of Physics,\\
Chongqing University, Chongqing 401331, China}
\author[\symknight]{~Ning Chen}
\emailAdd{chenning@ustc.edu.cn}
\affiliation[\symknight]{Department of Modern Physics, \\
University of Science and Technology of China, Hefei, Anhui 230026, China}

\date{\today}

\abstract
{\\[1mm]
In this work, we study the SM-like Higgs pair productions in the framework of the general CP-violating two-Higgs-doublet model.
Several constraints are imposed to the model sequentially, including the SM-like Higgs boson signal fits, the precise measurements of the electric dipole moments, the perturbative unitarity and stability bounds to the Higgs potential, and the most recent LHC searches for the heavy Higgs bosons.
We show how the CP-violating mixing angles are related to the Higgs cubic self couplings in this setup.
Based on these constraints, we suggest benchmark models for the future high-energy collider searches for the Higgs pair productions.
The $e^+ e^-$ colliders operating at $\sqrt{s}= (500\,\GeV\,, 1\,\TeV)$ are capable of measuring the Higgs cubic self couplings of the benchmark models directly.
Afterwards, we estimate the cross sections of the resonance contributions to the Higgs pair productions for the benchmark models at the future LHC and SppC/Fcc-hh runs.
Other possible decay modes for the heavy Higgs bosons are also discussed.
}

\arxivnumber{arXiv:1607.02703}

\begin{document}
\maketitle

\setcounter{page}{2}

\tableofcontents



\section{Introduction}
\label{section:intro}

The discovery of the $125\,\GeV$ Higgs boson~\cite{Aad:2012tfa, Chatrchyan:2012xdj} at the LHC runs at $7\oplus 8\,\TeV$ validate Higgs mechanism for the spontaneous breaking of the electroweak gauge symmetry (EWSB). 
The current LHC measurements of the Higgs boson couplings to the SM fermions, gauge bosons, and loop-induced couplings to photons and gluons reach the precision of $\sim 10-20\,\%$ level.
Besides, it is important to probe the Higgs self couplings to confirm the mechanism of the EWSB.
This can be done by looking for the Higgs pair productions at both high-energy $e^+ e^- $ and $pp$ colliders.
The current LHC searches for the Higgs pair productions focus on the leading production channel of gluon-gluon fusion (ggF), as well as the promising final states of $b \bar b \gamma\gamma$.
Some of the detailed studies at the LHC can be found in Refs.~\cite{Baur:2002rb,Baur:2002qd,Baur:2003gpa,Contino:2012xk, Dolan:2012rv, Papaefstathiou:2012qe, Baglio:2012np,Barger:2013jfa,Chen:2013emb, Chen:2014xra,Cao:2015oaa}.
From the experimental side, it is well-known that several future high-energy collider programs, such as the International Linear Collider (ILC)~\cite{Baer:2013cma} in Japan, the Future eplus-eminus/hadron-hadron Cicular Collider (Fcc-ee/Fcc-hh)~\cite{Gomez-Ceballos:2013zzn} at CERN, and the Circular electron-positron Collider (CEPC)/ Super-$pp$-Collider(SppC)~\cite{CEPC-SPPC-pre} in China, have been proposed in recent years.
A key physical goal for these different high-energy collider programs is try to probe the shape of the Higgs potential.
Some of the recent studies of the Higgs pair searches at the future colliders can be found in Refs.~\cite{Tian:2010np, Yao:2013ika,Hespel:2014sla, Barr:2014sga,Papaefstathiou:2015iba, Kotwal:2015rba, He:2015spf,Arhrib:2015hoa,Fuks:2015hna,Arkani-Hamed:2015vfh,Baglio:2015wcg,Hashemi:2015swh,Contino:2016spe}.

In many of new physics models beyond the SM (BSM), the Higgs sector is extended with several scalar multiplets. 
The two-Higgs-doublet model (2HDM) is one attractive alternative to the SM, which allows for new phenomena in the scalar sector~\cite{Branco:2011iw}.
To discover another Higgs doublet in the future LHC experiments, a lot of efforts have been taken for the $t \bar t$ searches with the novel techniques~\cite{Craig:2015jba, Hajer:2015gka,Chen:2015fca,Craig:2016ygr}, the decays of $hZ$ final states~\cite{Chen:2014dma}, and the charged Higgs searches~\cite{Yang:2011jk} as well.~\footnote{See also Refs.~\cite{Craig:2013hca,Djouadi:2015jea} for recent summaries of various search modes in the 2HDM at the LHC 14 TeV experiments.}
Most of the current studies focus on the CP-conserving (CPC) version of 2HDM.
Originally, the 2HDM was motivated to offer extra CP-violation (CPV) sources from the scalar sector~\cite{Lee:1974jb}. 
Recently, it was also pointed out that the CPV 2HDM is likely to realize the EW baryogenesis~\cite{Bian:2014zka}, which is one of the most popular solutions to the baryon asymmetry in the Universe.
Three neutral Higgs bosons, denoted as $(h_1\,, h_2\,, h_3)$, mix with each other in the CPV 2HDM.
There are two angles of $\alpha_b$ and $\alpha_c$ to parametrize the size of the CPV effects, and the CPC limit can be easily restored by taking $\alpha_b=\alpha_c=0$.
The $125\,\GeV$ SM-like Higgs boson, often chosen to be $h_1$ in the spectrum, is a mixture of both CP-even and CP-odd states~\cite{Lavoura:1994fv, Barroso:2012wz, Brod:2013cka,Mao:2014oya,Mao:2016jor}.
Such CPV couplings for the SM-like Higgs bosons are subject to the constraints from the searches for the electric dipole moments (EDMs) of the neutron, atoms, and molecules.~\footnote{See, e.g., Refs.~\cite{Pospelov:2005pr,Engel:2013lsa} for recent reviews. }
One of the most stringent one is from the ACME collaboration~\cite{Baron:2013eja}, where they reported an upper limit on the electron EDM (eEDM) of $|d_e/e|<8.7\times 10^{-29}\,{\rm cm} $.
This bound can be translated to constrain the size of the CPV mixing through the Barr-Zee type diagrams. 
More specifically, we find that the sizes of the CPV mixings also determine the sizes of the Higgs cubic self couplings.
Together with other existing constraints to the CPV 2HDM, which include the $125\,\GeV$ Higgs boson signal strengths, the perturbative unitarity and stability of the Higgs potential, and the constraints from the LHC searches for the heavy Higgs bosons, one can find the constraints to the heavy Higgs boson mass ranges and the sizes of the Higgs cubic self couplings.
Therefore, the cross sections of the Higgs pair productions in the CPV 2HDM can be envisioned for the future experimental searches at the LHC and the SppC.

This paper aims to study the Higgs pair productions in the framework of the CPV 2HDM, including the precise measurement of the SM-like Higgs cubic self couplings at the $e^+ e^-$ colliders, and the resonance contributions in the gluon-gluon fusion (ggF) production channel at the $pp$ colliders.
The layout of this paper is described as follows. 
In Sec.~\ref{section:CPV2HDM}, we review the setup of the CPV 2HDM.
With the assumptions of the degenerate heavy Higgs boson mass spectrum, we take the simplified parameter sets of $\alpha=-\pi/4$.
We also obtain the gauge couplings, Yukawa couplings, and the self couplings for Higgs bosons in the physical basis. 
In Sec.~\ref{section:constraint}, we impose series of constraints to the CPV 2HDM parameter space.
The combined constraints of 125 GeV Higgs signals and the eEDM bounds point to the $t_\beta\sim 1$ parameter choice. 
The size of the CPV mixing angle $|\alpha_b|$ is also bounded from above. 
For the CPV 2HDM-I, the CPV mixing is stringently constrained to be $|\alpha_b| \lesssim 5\times 10^{-3}$, which is quite approaching to the CPC limit. 
For the CPV 2HDM-II, the constraints to the CPV mixing are much relaxed, and we focus on this case for the Higgs pair productions.
The constraints from the unitarity, the stability, and the current LHC 8 TeV searches for the heavy Higgs bosons further restrict the allowed mass ranges of the heavy Higgs bosons and the soft $\mathbb{Z}_2$-breaking mass term of $m_{\rm soft}$. 
The main results of the Higgs pair productions in the CPV 2HDM are presented in Sec.~\ref{section:collider}.
By combining the current constraints, we show that the variations of the Higgs cubic self couplings are controlled by the size of the CPV mixing angle $|\alpha_b|$ and the soft mass term $m_{\rm soft}$ in the 2HDM potential.
A set of benchmark models are given with the fixed CPV mixing angles and the maximally allowed soft mass terms.
Under the small CPV limit, the Higgs cubic self coupling of $\lambda_{111}$ for the SM-like Higgs boson tends to the SM predicted value of $\lambda_{hhh}^{\rm SM}\simeq 32\,\GeV$, and the resonance contributions become negligible as well.
The corresponding Higgs pair production cross sections will tend to the predictions for the SM case.
We estimate the physical opportunities of the precise measurement of the SM-like Higgs cubic self coupling $\lambda_{111}$ at the future high-energy $e^+ e^-$ colliders, with focus on the $e^+ e^- \to hhZ$ process at the $\sqrt{s}=500\,\GeV$ run.
On the other hand, the heavy resonance contributions to the Higgs pair productions can become dominant at the $pp$ colliders.
The cross sections for the possible experimental search modes of $h_1 h_1\to (b \bar b \gamma\gamma\,, b \bar b WW)$ are estimated for both LHC $14\,\TeV$ and SppC/Fcc-hh $100\,\TeV$ runs.
In addition, several other possible search modes of $(W^+ W^-\,,ZZ\,, hZ)$ are also mentioned.
The conclusions and discussions are given in Sec.~\ref{section:conclusion}.


\section{The CPV 2HDM}
\label{section:CPV2HDM}

\subsection{The CPV 2HDM potential}

In the general 2HDM, two Higgs doublets of $(\Phi_1\,, \Phi_2)\in 2_{+1}$ are introduced in the scalar sector. 
For simplicity, we consider the soft breaking of a discrete $\mathbb{Z}_{2}$ symmetry, under which two Higgs doublets transform as $(\Phi_1\,,\Phi_2)\to (-\Phi_1\,, \Phi_2)$. 
The corresponding Lagrangian is expressed as
\beqs
\beqn
\mL&=&\sum_{i=1\,,2}|D \Phi_i|^2 - V(\Phi_1\,,\Phi_2)\,,\\
V(\Phi_1\,,\Phi_2)&=&m_{11}^2|\Phi_1|^2+m_{22}^2|\Phi_2|^2-(m_{12}^2 \Phi_1^\dag\Phi_2+H.c.)+\hf\lambda_1 |\Phi_{1}|^{4} +\hf\lambda_2|\Phi_{2}|^{4}\non
&+&\lambda_3|\Phi_1|^2 |\Phi_2|^2+\lambda_4 |\Phi_1^\dag \Phi_2|^2+\hf \Big[ \lambda_5 (\Phi_1^\dag\Phi_2)^2 +H.c.\Big]\,,\label{eq:2HDM_potential}
\eeqn
\eeqs
with $(m_{12}^2\,, \lambda_5)$ being complex and all other parameters being real for the CPV 2HDM.
After the EWSB, two Higgs doublets $\Phi_1$ and $\Phi_2$ in the unitarity gauge can be expressed as
\beqn\label{eq:2HDM_doublets}
&&\Phi_1=\left(  
\ba{c}  -s_\beta\, H^+ \\
 \frac{1}{\sqrt{2}} ( v_1 + H_1^0 - i s_\beta A^0)  
 \ea  \right)\,,\qquad 
 \Phi_2= \left(  
\ba{c}  c_\beta\, H^+ \\
 \frac{1}{\sqrt{2}} ( v_2 e^{i\xi} + H_2^0 + i c_\beta A^0 ) 
 \ea  \right)\,,
\eeqn
where $v_1^2+v_2^2 = v^2 = (\sqrt{2}\, G_F)^{-1}$.
The ratio between two Higgs VEVs is parametrized as
\beqn\label{eq:tb}
t_\beta&\equiv&\tan\beta = \frac{v_2}{v_1}\,,
\eeqn
and $\xi$ represents the relative phase between two Higgs doublets.
The imaginary components of $m_{12}^2$ and $\lambda_5$ are the source of CP violation, which lead to the mixings among three neutral states as $(h_1\,, h_2\,, h_3)^T=\mR\,(H_1^0\,, H_2^0\,, A^0)^T$.
Explicitly, the $3\times 3$ mixing matrix is expressed as~\cite{Khater:2003wq}
\beqn\label{eq:neutral_mix}
\mR&=&\mR_{23}(\alpha_c) \mR_{13}(\alpha_b) \mR_{12}(\alpha+\frac{\pi}{2})\non
&=&\left( \ba{ccc} 
-s_\alpha c_{\alpha_b} & c_\alpha c_{\alpha_b} & s_{\alpha_b} \\
  s_\alpha s_{\alpha_b}s_{\alpha_c} - c_\alpha c_{\alpha_c} & -s_\alpha c_{\alpha_c} - c_\alpha s_{\alpha_b} s_{\alpha_c} & c_{\alpha_b}s_{\alpha_c}  \\
  s_\alpha s_{\alpha_b} c_{\alpha_c} + c_\alpha s_{\alpha_c}  &  s_\alpha s_{\alpha_c} - c_\alpha s_{\alpha_b} c_{\alpha_c}  &  c_{\alpha_b} c_{\alpha_c}  \\ 
   \ea  \right)\,.
\eeqn
The angle $\alpha$ parametrizes the mixing between two CP-even states of $(H_1^0\,,H_2^0)$.
The CPV mixing angles of $\alpha_b$ and $\alpha_c$ parametrize the CP mixings between $(H_1^0\,,A^0)$ and $(H_2^0\,,A^0)$, respectively. 
Their ranges are taken as
\beqn
&&- \frac{\pi}{2} \leq \alpha_b \leq \frac{\pi}{2}\,,\qquad  - \frac{\pi}{2} \leq \alpha_c \leq \frac{\pi}{2}\,.
\eeqn
In the CPC limit, one has $\alpha_b = \alpha_c =0$. 
Correspondingly, $\mR$ becomes block diagonal, and $(h_1\,,h_2)$ are purely CP-even states.

By minimizing the CPV 2HDM potential, one obtains the following relations for the mass parameters
\beqs\label{eqs:2HDM_mini}
\beqn
m_{11}^2 &=& {\rm Re} (m_{12}^2 e^{i\xi}) t_\beta -\frac{1}{2} \Big[ \lambda_1 v^2 c_\beta^2 + (\lambda_3 + \lambda_4) v^2 s_\beta^2 + {\rm Re} (\lambda_5 e^{2i\xi}) v^2 s_\beta^2 \Big] \ , \hspace{0.8cm}
\label{eq:2HDM_mini1}\\
m_{22}^2 &=&{\rm Re} (m_{12}^2 e^{i\xi})/t_\beta  - \frac{1}{2}\Big[ \lambda_2 v^2 s_\beta^2 + (\lambda_3 + \lambda_4) v^2 c_\beta^2 + {\rm Re} (\lambda_5 e^{2i\xi}) v^2 c_\beta^2 \Big] \ ,
\label{eq:2HDM_mini2}\\
{\rm Im} (m_{12}^2 e^{i\xi} )&=&\hf v^2 s_\beta c_\beta {\rm Im} ( \lambda_5  e^{2i\xi } ) \,.\label{eq:2HDM_mini3}
\eeqn
\eeqs
The physical masses of $(M_1\,, M_2\,, M_3\,, M_\pm)$ in the scalar spectrum are obtained from the 2HDM potential together with the minimization conditions given in Eqs.~\eqref{eqs:2HDM_mini}.
The charged Higgs boson mass squared reads
\beqn
M_\pm^2&=& \frac{1}{s_\beta c_\beta} {\rm Re}(m_{12}^2 e^{i\xi} ) - \hf \Big[ \lambda_4 + {\rm Re}(\lambda_5 e^{2i\xi} )  \Big] v^2\,.
\eeqn
The mass squared matrix for the neutral sector can be expressed as
\beqn\label{eq:neutral_mass_matrix}
\mM_0^2&=& \left(  
\ba{ccc}     
\lambda_1 c_\beta^2 + \nu s_\beta^2 &  ( \lambda_{345} -\nu ) s_\beta c_\beta  &  -\hf {\rm Im}( \lambda_5e^{2i\xi} ) \, s_\beta   \\
  ( \lambda_{345} -\nu ) s_\beta c_\beta  &    \lambda_2 s_\beta^2 + \nu c_\beta^2   &    -\hf {\rm Im}( \lambda_5 e^{2i\xi} ) \, c_\beta    \\
  - \hf {\rm Im}( \lambda_5 e^{2i\xi} ) \, s_\beta &  -\hf {\rm Im}(\lambda_5 e^{2i\xi}) \, c_\beta  & -{\rm Re} ( \lambda_5 e^{2i\xi} ) + \nu  \\
\ea  \right) v^2\,,
\eeqn
with the short-handed notations of
\beqn\label{eq:nu}
&& \lambda_{345}\equiv \lambda_3+\lambda_4+ {\rm Re}(\lambda_5 e^{2i\xi} )\,,\qquad \nu \equiv \frac{ {\rm  Re}(m_{12}^2 e^{i\xi} )}{v^2 s_\beta c_\beta }\,.
\eeqn
By diagonalizing the mass squared matrix with the mixing matrix in Eq.~\eqref{eq:neutral_mix}, one has
\beqn
\mM_0^2&=& \mR^T \, {\rm diag}(M_1^2\,, M_2^2\,, M_3^2 )\, \mR\,,
\eeqn
from which one further obtains the relations to trade the quartic Higgs self couplings into the physical inputs as follows
\beqs\label{eqs:lambdas} 
\beqn
\lambda_1 &=& \frac{ M_1^2 \mR_{11}^2 + M_2^2 \mR_{21}^2 + M_3^2 \mR_{31}^2}{v^2 c_\beta^2} - \nu\, t_\beta^2 \ , \\
\lambda_2 &=& \frac{ M_1^2 \mR_{12}^2 + M_2^2 \mR_{22}^2  + M_3^2 \mR_{32}^2}{v^2 s_\beta^2} - \nu\, /t_\beta^2 \ , \\
\lambda_3&=&- \nu + \frac{2 M_\pm^2 }{v^2}  + \frac{ M_1^2 \mR_{11}\mR_{12} + M_2^2 \mR_{21} \mR_{22} + M_3^2 \mR_{31} \mR_{32}  }{v^2 s_\beta c_\beta } \,,\\
\lambda_4 &=& 2\nu - \frac{2 M_\pm^2 }{v^2} -{\rm Re}( \lambda_5e^{2i\xi} )  \,, \\
{\rm Re}( \lambda_5e^{2i\xi} ) &=& \nu - \frac{ M_1^2 \mR_{13}^2 + M_2^2 \mR_{23}^2 + M_3^2 \mR_{33}^2 }{v^2} \ , \\
{\rm Im}(\lambda_5 e^{2i\xi} ) &=& -\frac{1}{v^2 s_\beta c_\beta} \Big[  ( M_1^2 \mR_{11} \mR_{13} + M_2^2 \mR_{21} \mR_{23} + M_3^2 \mR_{31} \mR_{33} )c_\beta\non
& +& ( M_1^2 \mR_{12} \mR_{13} + M_2^2 \mR_{22} \mR_{23} + M_3^2 \mR_{32} \mR_{33} )  s_\beta  \Big] \ . \label{eq:imlambda5} 
\eeqn
\eeqs
For simplicity, we can always work in the basis where $\xi=0$ by using the rephasing invariance. 
We also assume that ${\rm  Re}(m_{12}^2 ) \ge 0$, and use the notation for the soft mass term as
\beqn\label{eq:msoft}
&& m_{\rm soft}^2\equiv{\rm  Re}(m_{12}^2 )\,.
\eeqn
The elements of $(\mM_0^2)_{13}$ and $(\mM_0^2)_{23}$ in Eq.~\eqref{eq:neutral_mass_matrix} provide the CPV mixings, which are related via $t_\beta$ as
\beqn
&&(\mM_0^2)_{13} = (\mM_0^2)_{23} \, t_\beta\,.
\eeqn
This leads to one additional constraint between mixing angles and mass eigenvalues as follows~\cite{Khater:2003wq}
\beqn\label{eq:mass_constraint}
&&( M_1^2 - M_2^2 s_{\alpha_c}^2  - M_3^2 c_{\alpha_c}^2 ) s_{\alpha_b} (1 + t_\alpha)= (M_2^2 - M_3^2 ) (t_\alpha t_\beta - 1) s_{\alpha_c} c_{\alpha_c}\,.
\eeqn
In the analysis below, we always identify $h_1$ as the SM-like Higgs boson with mass of $125\,\GeV$.
We further simplify the parameter inputs by requiring all heavy Higgs boson masses are degenerate, i.e., $M_{2}=M_{3}=M_\pm \equiv M$. 
This was usually taken to relax the constraints from the electroweak precision measurements. 
The constraint of Eq.~\eqref{eq:mass_constraint} among the mixing angles becomes
\beqn
&&\alpha_b=0 \,,\qquad \textrm{or}~~~ t_\alpha=-1\,.
\eeqn
Below, we will always take $\alpha=-\pi/4$.~\footnote{
The study of the phenomenology with the CPV mixings of $|\alpha_b|\ll |\alpha_c|$ is carried out in a separate work~\cite{Bian:2016zba}.
}
The input parameters of $(\beta\,, \alpha_b)$ will be determined through other constraints.
Since $\alpha_c$ determines the size of the CPV mixing between two mass-degenerate Higgs bosons of $h_2$ and $h_3$ in our setup, one can anticipate that $\alpha_c$ becomes unphysical in physical processes to be studied below.
Without loss of generality, we always take $\alpha_c=0$ for simplicity.

Thus, the set of input parameters can be summarized as follows
\beqn\label{eq:inputs}
&&M_1=125\,\GeV\,, \qquad M_2=M_3=M_\pm = M\,, \qquad m_{\rm soft} \non
&&\alpha=-\frac{\pi}{4}\,,\qquad  t_\beta \,, \qquad  \alpha_b\,,\qquad \alpha_c=0 \,.
\eeqn
Analogous to the CPC version of the general 2HDM, the parameter choice of $\beta-\alpha=\pi/2$ corresponds to the so-called ``alignment limit''.
This can be achieved when taking into account the signal fit to the 125 GeV SM-like Higgs boson $h_1$, as shown later.
By further combining with the eEDM constraints, we will fix the parameters of $t_\beta$ and $\alpha_b$ and constrain two other mass parameters of $M$ and $m_{\rm soft}$ for our later discussions.

\subsection{The couplings in the CPV 2HDM}

For simplicity, we focus on the 2HDMs where the Yukawa sector has a $\mathbb{Z}_2$ symmetry and $\Phi_1$ and $\Phi_2$ each only gives mass to up-type quarks or down-type quarks and charged leptons. 
This is sufficient to suppress tree-level flavor changing processes mediated by the neutral Higgs bosons.
The Yukawa couplings for the 2HDM-I and 2HDM-II read (and suppressing the CKM mixing),
\beqn\label{eq:2HDM_Yuk}
\mathcal{L}= \left\{\begin{array}{ll} 
-\biggl( \displaystyle{ c_\alpha\over s_\beta}{m_u\over v} \biggr)\overline Q_L \tilde \Phi_2 u_R  -\biggl(  {  c_\alpha\over s_\beta}{m_d\over v}\biggr) \overline Q_L \Phi_2 d_R + {\rm h.c.} & \hspace{1cm} {\rm 2HDM-I}\vspace{0.2cm} \\
-\biggl(  \displaystyle{ c_\alpha\over s_\beta}{m_u\over v} \biggr)\overline Q_L \tilde \Phi_2 u_R 
+\biggl( { s_\alpha\over c_\beta}{m_d\over v} \biggr)\overline Q_L \Phi_1 d_R
+ {\rm h.c.} & \hspace{1cm} {\rm 2HDM-II} \, ,
\end{array} \right.
\eeqn
where $Q_L^T=(u_L,d_L)$ and $\tilde \Phi_2 \equiv i \sigma_2 \Phi_2^*$.
For both cases, the charged lepton Yukawa coupling has the same form as that of the down-type quarks.
Therefore, we can express the couplings between neutral Higgs bosons and the fermions and gauge bosons in the mass eigenbasis
\beqn
\mL&=& \sum_{i=1}^3 \left[-m_f\left( c_{f,i} \bar f f+ \tilde c_{f,i} \bar f i\gamma_5 f  \right) + a_i \left( 2  m_W^2 W_\mu W^\mu +  m_Z^2 Z_\mu Z^\mu \right)  \right] \frac{h_i}{v}  \ .
\label{coup_f}
\eeqn
When $c_{f,i}\tilde c_{f,i}\neq 0$ or $a_{i}\tilde c_{f,i}\neq 0$, the mass eigenstate $h_i$ couples to both CP-even and CP-odd operators, so the CP symmetry is violated.
The coefficients of $c_{f,i}$, $\tilde c_{f,i}$ and $a_i$ can be derived from the elements of the rotation matrix $\mR$ defined in Eq.~\eqref{eq:neutral_mix}, which were also previously obtained in Refs.~\cite{Shu:2013uua, Inoue:2014nva, Chen:2015gaa}. 
Here, we summarize their explicit expressions under the alignment limit in Table.~\ref{tab:Hcouplings}.
In this alignment limit of $\beta-\alpha=\pi/2$, the Higgs Yukawa couplings and Higgs gauge couplings are determined by the CPV mixing angles of $(\alpha_b\,, \alpha_c)$ and $t_\beta$.
By taking the CPC limit of $\alpha_b=\alpha_c=0$, it is evident that $(h_1\,, h_2)$ have the purely CP-even Yukawa couplings of $c_{f\,,i}$, while $h_3$ has the purely CP-odd Yukawa couplings of $\tilde c_{f\,,i} $.
The previous studies of the collider measurements of the CPV in the Higgs Yukawa couplings can be found in Refs.~\cite{Harnik:2013aja,Berge:2013jra,Brod:2013cka,Askew:2015mda,Li:2015kxc,Buckley:2015vsa,Berge:2015nua,Hagiwara:2016rdv,Rindani:2016scj}.

\begin{table}[htb]
\begin{center}
\begin{tabular}{c|c|c}
\hline
\hline
 & 2HDM-I & 2HDM-II   \\\hline
 $c_{u\,,1}$  &  $ c_{\alpha_b}$  & $ c_{\alpha_b}$  \\
 $c_{d\,,1}=c_{\ell\,,1}$  & $ c_{\alpha_b}$     &  $c_{\alpha_b}$   \\
 $\tilde c_{u\,,1}$  & $-s_{\alpha_b}/t_\beta$   &  $-s_{\alpha_b}/t_\beta$    \\
 $\tilde c_{d\,,1}=\tilde c_{\ell\,,1}$  &  $s_{\alpha_b}/t_\beta$   &  $-s_{\alpha_b}\,t_\beta$   \\
 $a_1$  &  $ c_{\alpha_b}$  &   $ c_{\alpha_b}$  \\
 \hline
 $c_{u\,,2}$  & $  c_{\alpha_c}/t_\beta - s_{\alpha_b} s_{\alpha_c}$   &  $ c_{\alpha_c}/t_\beta - s_{\alpha_b} s_{\alpha_c}$   \\
 $c_{d\,,2}=c_{\ell\,,2}$  &  $ c_{\alpha_c}/t_\beta - s_{\alpha_b} s_{\alpha_c}$   & $- s_{\alpha_b} s_{\alpha_c} - c_{\alpha_c}\,t_\beta  $    \\
 $\tilde c_{u\,,2}$  &  $-c_{\alpha_b}s_{\alpha_c}/t_\beta$   &   $-c_{\alpha_b}s_{\alpha_c}/t_\beta$  \\
 $\tilde c_{d\,,2}=\tilde c_{\ell\,,2}$  &  $c_{\alpha_b}s_{\alpha_c}/t_\beta$   &  $-c_{\alpha_b}\,s_{\alpha_c}\,t_\beta$   \\
 $a_2$  &  $-s_{\alpha_b} s_{\alpha_c}  $  & $-  s_{\alpha_b} s_{\alpha_c} $  \\
 \hline
 $c_{u\,,3}$  & $ - s_{\alpha_c}/t_\beta  -  s_{\alpha_b} c_{\alpha_c} $   &  $ - s_{\alpha_c}/t_\beta  -  s_{\alpha_b} c_{\alpha_c} $   \\
 $c_{d\,,3}=c_{\ell\,,3}$  &  $ - s_{\alpha_c}/t_\beta  -  s_{\alpha_b} c_{\alpha_c}$   &  $- s_{\alpha_b} c_{\alpha_c}+ s_{\alpha_c}\, t_\beta $   \\
 $\tilde c_{u\,,3}$  &  $-c_{\alpha_b}c_{\alpha_c}/t_\beta$  &  $-c_{\alpha_b}c_{\alpha_c}/t_\beta$   \\
 $\tilde c_{d\,,3}=\tilde c_{\ell\,,3}$  &  $c_{\alpha_b}c_{\alpha_c}/t_\beta$   &  $-c_{\alpha_b}c_{\alpha_c}\, t_\beta$   \\
 $a_3$  &   $ - s_{\alpha_b} c_{\alpha_c} $ &   $ -  s_{\alpha_b} c_{\alpha_c} $   \\
 \hline\hline
\end{tabular}
\caption{The SM fermion and gauge boson couplings to Higgs mass eigenstates in the alignment of $\beta-\alpha = \pi/2$. }\label{tab:Hcouplings}
\end{center}
\end{table}

By extracting the cubic terms in the scalar potential Eq.~\eqref{eq:2HDM_potential}, we can obtain the Higgs cubic self-interacting terms.
The neutral part of the cubic terms are expressed as follows in the basis of $(H_1^0, H_2^0, A^0)$
\beqn\label{eq:L3s}
-\mL_{3s}/v &=& \hf \lambda_1 c_\beta (H_1^0)^3 + \hf \lambda_2 s_\beta (H_2^0)^3 + \hf \lambda_{345} \Big[ c_\beta H_1^0 (H_2^0)^2 + s_\beta H_2^0 (H_1^0)^2  \Big]\non
&+& \hf \Big\{ c_\beta \Big[ \lambda_1 s_\beta^2 + \lambda_{345} c_\beta^2 -2 {\rm Re}(\lambda_5)  \Big] H_1^0 +  s_\beta \Big[ \lambda_2 c_\beta^2 + \lambda_{345} s_\beta^2 -2 {\rm Re}(\lambda_5)   \Big] H_2^0 \Big\} (A^0)^2\non
&-&  \hf {\rm Im}(\lambda_5) \Big\{ 2\,H_1^0 H_2^0 A^0 + s_\beta c_\beta \Big[ (H_1^0)^2 + (H_2^0)^2 - (A^0 )^2  \Big]  \Big\} A^0\,.
\eeqn
From these terms, one can readily obtain the cubic interactions in terms of the mass eigenstates of $(h_1, h_2, h_3)$ by using the orthogonal mixing matrix $\mR$ from Eq.~\eqref{eq:neutral_mix}. 
Throughout our discussions, we define the Higgs cubic self couplings of $\lambda_{ijk}\, (i\,,j\,,k=1\,,2\,,3)$ to be the coefficients of the $h_i h_j h_k$ term from Eq.~\eqref{eq:L3s}
\beqn\label{eq:lambda_ijk}
\lambda_{ijk}&\equiv&\frac{1}{S\,!}\frac{\partial^3 \mL_{3s}}{\partial h_i\, \partial h_j\, \partial h_k}\,,
\eeqn
where the symmetry factors are such that $S!=3!=6$ for $i=j=k$, $S!=2$ for $i=j\neq k$, and $S=1$ for $i\neq j \neq k$. 
A general derivation of the Higgs cubic self couplings in the CPV 2HDM was previously studied in Refs.~\cite{Pilaftsis:1999qt,Carena:2002bb,Osland:2008aw}.
The explicit expressions of $\lambda_{ijk}$ are tedious, while they can be greatly simplified with the fixed parameters through the following discussions.


\section{The Constraints in The CPV 2HDM}
\label{section:constraint}

\subsection{The $125\,\GeV$ SM-like Higgs boson constraint}

In the CPV 2HDM, the productions and decay rates of the 125 GeV SM-like Higgs boson $h_1$ are controlled by both CP-even couplings of $c_{f\,,1}$ and CP-odd couplings of $\tilde c_{f\,,1}$.
The production cross sections and decay rates are rescaled from the SM one as follows,
\beqs
\beqn
&&\frac{\sigma[gg\to h_1]}{\sigma[gg\to h_{\rm SM} ]}\approx \frac{(1.03\, c_{u\,,1} - 0.06\, c_{d\,,1})^2 +(1.57\, \tilde c_{u\,,1} - 0.06\, \tilde c_{d\,,1})^2 }{(1.03-0.06)^2}\,,\\
&& \frac{\Gamma[h_1\to \gamma\gamma]}{ \Gamma[h_{\rm SM}\to \gamma\gamma] } \approx \frac{ (0.23\, c_{u\,,1} - 1.04 \, a_1 )^2 + (0.35\, \tilde c_{u\,,1} )^2  }{ (0.23-1.04)^2}\,, \\
&& \frac{\sigma[VV\to h_1]}{\sigma[VV\to h_{\rm SM}] } = \frac{\sigma[V^* \to V h_1] }{\sigma[V^*\to V h_{\rm SM} ]} = \frac{\Gamma[h_1 \to VV^* ] }{ \Gamma[h_{\rm SM}\to VV^* ]  } = a_1^2 \,,\\
&&\frac{\Gamma[h_1\to b \bar b]}{ \Gamma[h_{\rm SM}\to b  \bar b] } = \frac{\Gamma[h_1\to\tau \tau]}{ \Gamma[h_{\rm SM}\to\tau \tau ] } \approx c_{d\,,1}^2 + \tilde c_{d\,,1}^2\,.
\eeqn
\eeqs
For the production cross sections and decay rates of the SM Higgs boson, we use the results from the LHC Higgs Working Group given in Refs.~\cite{Dittmaier:2011ti, Denner:2011mq}.
The LHC signal strengths of the SM-like Higgs boson in the presence of the CPV were discussed in Refs.~\cite{Shu:2013uua, Inoue:2014nva, Chen:2015gaa, Freitas:2012kw, Celis:2013rcs, Djouadi:2013qya, Chang:2013cia, Fontes:2014tga,Fontes:2014xva,Fontes:2015mea,Fontes:2015xva}.
From Table.~\ref{tab:Hcouplings}, one notes that the relevant Yukawa couplings of $(c_{f\,,1}\,, \tilde c_{f\,,1})$ and the Higgs gauge couplings of $a_1$ are only controlled by the Higgs VEV ratio of $t_\beta$ as well as the CPV mixing angle of $\alpha_b$.
The heavy Higgs bosons in the spectrum are either irrelevant or negligible for the signal fit of $h_1$.
Based on the most recent LHC measurements of the $125\,\GeV$ signal strengths~\cite{TheATLAScollaboration:2013lia,ATLAS:2014yka,Aad:2014eha,ATLAS:2014aga, Khachatryan:2014jba,Aad:2015vsa}, we fit the signal strength of $h_1$ on the $(t_\beta\,, |\alpha_b|)$ plane and present the results with the eEDM constraints later.

\subsection{The eEDM constraints}

The ACME experiment~\cite{Baron:2013eja},which searches for an energy shift of ThO molecules due to an external electric field, set stringent experimental bound to the eEDM.~\footnote{As noted by~\cite{Dekens:2014jka} that current limits on the hadronic EDMs might provide similar sensitivities as the electron EDM, roughly $d_e/d_n\sim10^{-2}$, thus one could expect that $^{199}$Hg measurement \cite{Khater:2003wq} would give rises to complementary constraints on CP phases though hadronic EDMs are subjected to 
uncertainties of hadronic matrix elements~\cite{Engel:2013lsa}. }
The bound reads
\beqn\label{eq:ACME}
&& \Big| \frac{d_e}{e}  \Big| < 8.7\times 10^{-29}\, {\rm cm}\,.
\eeqn
The eEDM constraints to the CPV 2HDM-II were previously studied in the Refs.~\cite{Inoue:2014nva, Bian:2014zka}.
The effective Lagrangian term is given as follows
\beqn\label{eq:eEDM_eff}
\mL_{\rm eff}&=& -\frac{i}{2} d_e \bar e \sigma_{\mu\nu} \gamma_5 e\, F^{\mu\nu} = i \frac{e \delta_e m_e }{v^2 }  \bar e \sigma_{\mu\nu} \gamma_5 e \, F^{\mu\nu}\,,
\eeqn
after integrating out the internal heavy degrees of freedoms.  
The constraint in Eq.~\eqref{eq:ACME} can be converted to the bounds of the dimensionless Wilson coefficient of $\delta_e$ in Eq.~\eqref{eq:eEDM_eff} such as
\beqn\label{eq:EDM_Wilson}
&& \frac{2 m_e}{ v^2 } |\delta_e | < 8.7 \times 10^{-29}\, {\rm cm}\,.
\eeqn
%

\begin{figure}[htb]
\centering
\includegraphics[height=4cm]{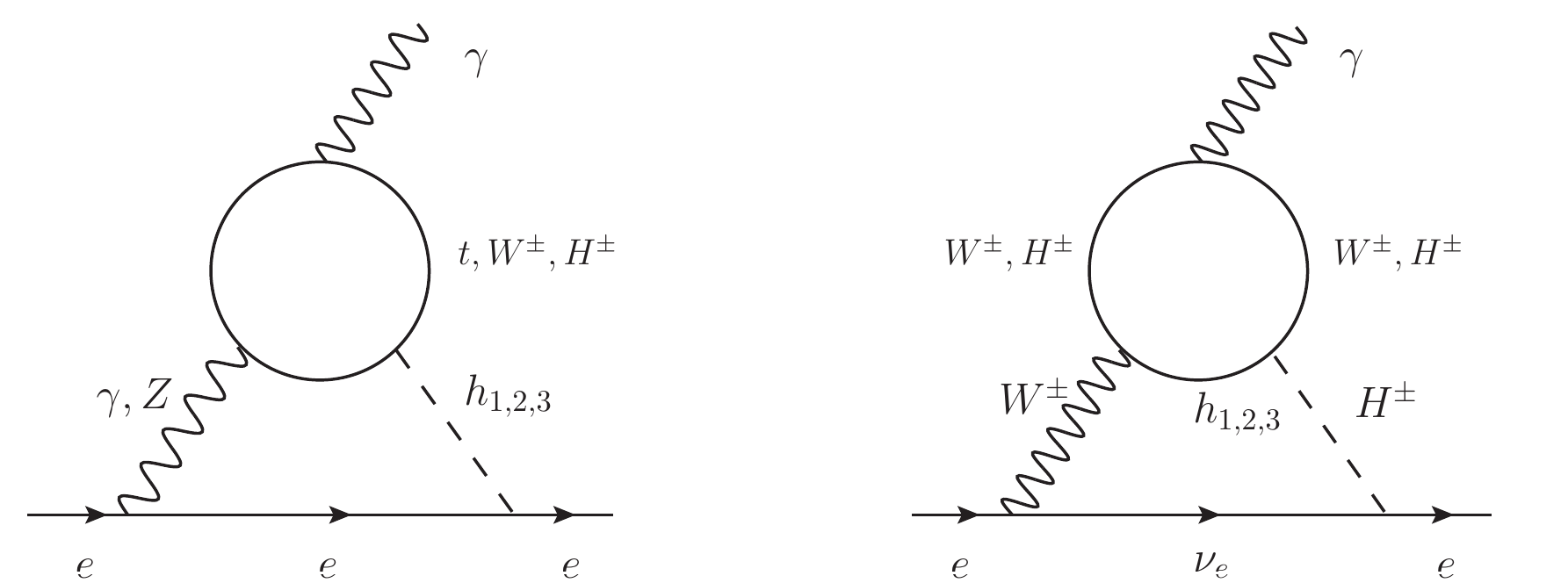}
\caption{\label{fig:BZ} 
Left: the eEDM from the Barr-Zee type diagrams with the $h_i V_{\mu\nu} V^{\mu\nu}$ or $h_i V_{\mu\nu} \tilde V^{\mu\nu}$ operators (with $V_{\mu\nu}= F_{\mu\nu}/Z_{\mu\nu}$), and the CPV couplings between the neutral Higgs bosons $h_i$ and the electron. 
Right: the eEDM from the $W^\pm H^\mp$ interactions and the CPV couplings for the charged Higgs bosons.
}
\end{figure}

\begin{figure}[htb]
\centering
\includegraphics[width=7.5cm,height=5cm]{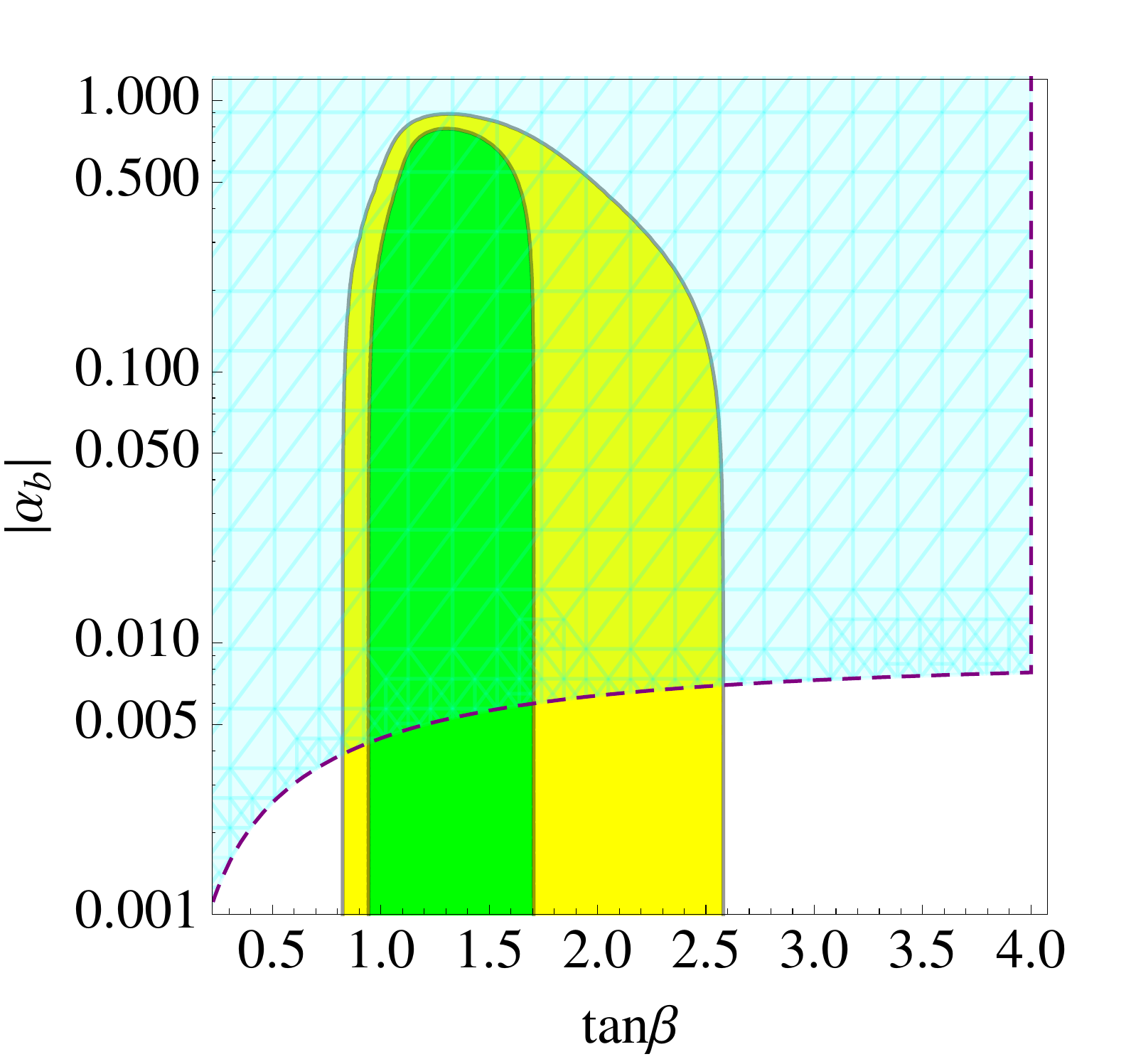}
\includegraphics[width=6.8cm,height=4.8cm]{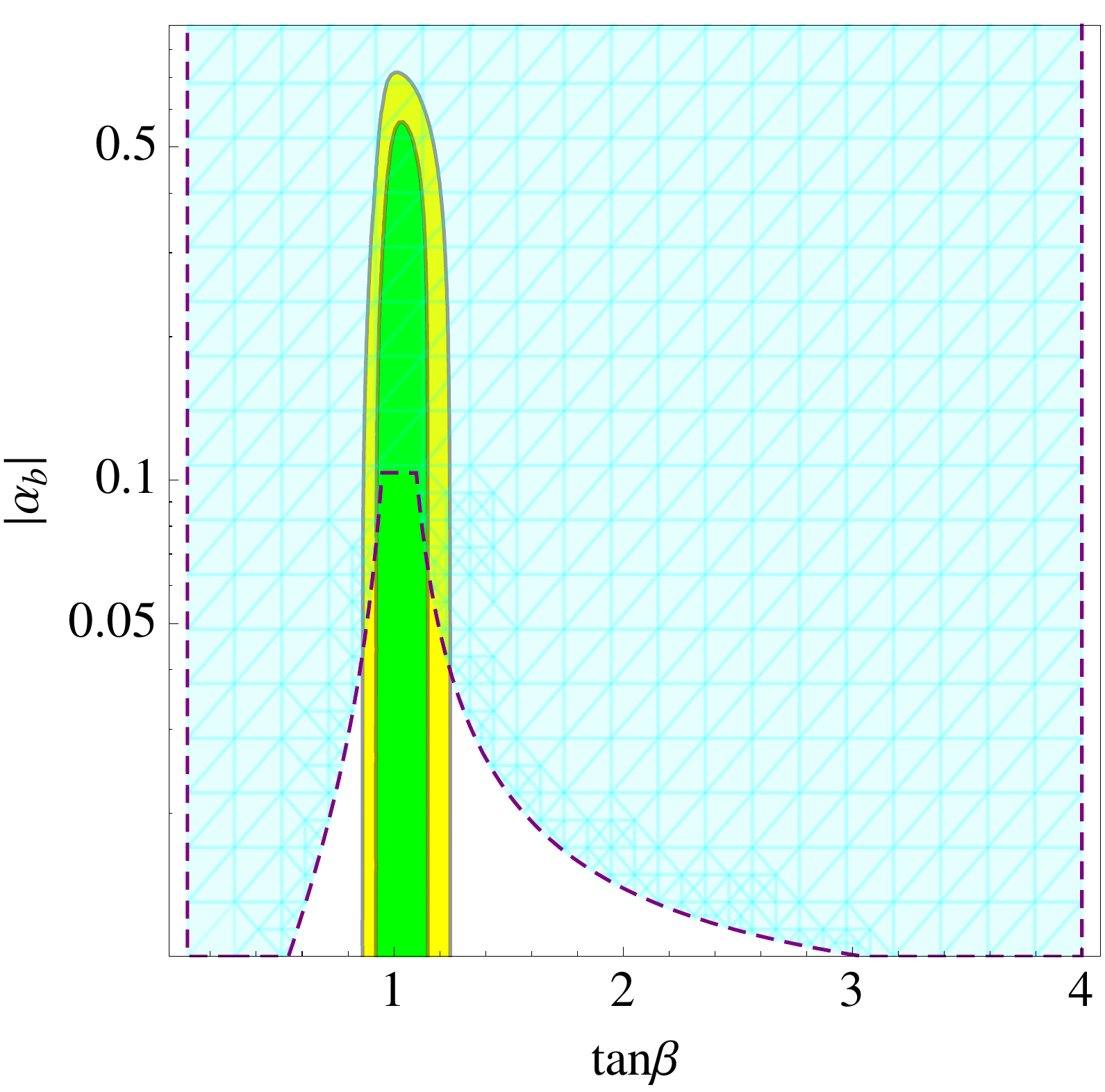}
\caption{\label{fig:h125sigEDM} 
The signal strength fit to the $125\,\GeV$ Higgs boson $h_1$ and the eEDM constraint (light-blue shaded region) on the $(t_\beta\,, |\alpha_b|)$ plane, left panel: CPV 2HDM-I, right panel: CPV 2HDM-II.
The green and yellow regions correspond to the $1\,\sigma$ and $2\,\sigma$ allowed regions for the LHC $7\oplus 8\,\TeV$ signal fit to the $h_1$ in the CPV 2HDM.
}
\end{figure}

In the CPV 2HDM, the Wilson coefficient $\delta_e$ are contributed by the two-loop Barr-Zee type $h_i \gamma\gamma$($h_i Z \gamma$) diagrams~\cite{Barr:1990vd}, and the $H^\pm W^\mp \gamma$ diagrams, as depicted in Fig.~\ref{fig:BZ}. 
The  $h_i \gamma\gamma$($h_i Z \gamma$) diagrams include the contributions from: (i) the top-quark loops, (ii) the $W$-boson and the NGB loops, and (iii) the charged Higgs boson loops. 
The total contributions can be summarized as follows
\beqn\label{eq:EDM_BZ}
\delta_e&=& (\delta_e)_t^{h_i \gamma\gamma} + (\delta_e)_W^{h_i \gamma \gamma}+ (\delta_e)_{H^\pm}^{h_i \gamma \gamma}\non
& +& (\delta_e)_t^{h_i Z \gamma}+ (\delta_e)_W^{h_i Z \gamma} + (\delta_e)_{H^\pm}^{h_i Z \gamma} +  (\delta_e)_{h_i}^{H^\pm W^\mp \gamma} \,.
\eeqn
Here, the superscripts of $h_i \gamma\gamma$, $h_i Z \gamma$, and $H^\pm W^\mp \gamma$ represent the operators for the specific Barr-Zee type diagrams.
The subscripts of $(t\,, W\,, H^\pm \,, h_i)$ represent the particles in the loops.
Explicit expression for each term can be found in Refs.~\cite{Chang:1990sf, Chang:1998uc, Abe:2013qla}, and summarized in the appendix of Ref.~\cite{Inoue:2014nva}.
Numerically, the leading contributions to the Wilson coefficient $\delta_e$ are mainly due to the $(\delta_e)^{h_1 \gamma\gamma}$ and $(\delta_e)^{h_1 Z \gamma}$ terms, while the contributions from the other heavy Higgs bosons of $(h_{2\,,3}\,, H^\pm)$ can be safely neglected.
These terms are proportional to the CP-odd couplings of $\tilde c_{f\,,1} $, and further proportional to the CPV mixing angle $\alpha_b$ according to the Yukawa couplings listed in Table.~\ref{tab:Hcouplings}.

The eEDM upper bound from the ACME is converted to the constraints to the CPV 2HDM parameters on the $(t_\beta\,, |\alpha_b|)$ plane. 
The combined $125\,\GeV$ Higgs boson signal constraints and the eEDM constraints are shown in Fig.~\ref{fig:h125sigEDM}.
It is clear that the eEDM bound is the leading one to set upper bounds to the CPV mixing angle of $|\alpha_b|$, as compared to the fits of the SM-like Higgs boson signal strengths.
For the CPV 2HDM-I (left panel), the size of CPV mixing angle is significantly bound as $|\alpha_b|\lesssim 5\times 10^{-3}$, and the $1\,\sigma$ allowed range of $t_\beta$ is within $(0.9\,,1.7)$.
For the CPV 2HDM-II (right panel), the allowed region of the CPV mixing angle can be extended to $|\alpha_b |\lesssim 0.1$, while the $1\,\sigma$ allowed range of $t_\beta$ is basically around $1.0$.
It has been noted in Ref.~\cite{Bian:2014zka} that the maximal cancellations between the $h_i F^{\mu\nu} V_{\mu\nu}$ operator and the $h_i F^{\mu\nu}\tilde V_{\mu\nu}$ operator can be achieved with the input of $t_\beta \sim 1$ in the CPV 2HDM-II.
In order to highlight the CPV effects in the Higgs self couplings in the following discussions, we will focus on the CPV 2HDM-II with the fixed inputs of $\alpha=-\pi/4$ and $t_\beta=1.0$.
Furthermore, we also find that the Higgs cubic self couplings almost approach to the SM limit when the CPV mixing angle can be constrained as small as $|\alpha_b|\lesssim 0.01$.
As stated in the previous paragraph, the Wilson coefficient of $\delta_e$ depends on the CPV mixings almost linearly.
Therefore, if the future measurements of the eEDM can improve the precisions to an order of magnitude or more, they can be very useful to constrain the benchmark models for the Higgs pair productions in this setup.

\subsection{The unitarity and stability constraints}

To have a self-consistent description of the 2HDM potential, two other theoretical constraints should be taken into account, namely, the perturbative unitarity and the stability.

Very roughly speaking, the perturbative unitarity constraint means that the theory cannot be strongly coupled.
According to the relations listed in Eqs.~\eqref{eqs:lambdas}, the constraints to the self couplings of $\lambda_i$ can be converted to upper bounds to the Higgs boson masses and the soft mass term of $m_{\rm soft}$ in the 2HDM.
In practice, the necessary and sufficient condition of the tree-level unitarity bounds can be obtained by evaluating the eigenvalues of the $S$-matrices for the scattering processes of the scalar fields in the 2HDM~\cite{Arhrib:2000is, Kanemura:2015ska}. 
Due to the Nambu-Goldstone theorem, the $S$-matrices can be expressed in terms of 2HDM quartic couplings $\lambda_i$. 
Explicitly, the unitarity conditions to be satisfied are that the eigenvalues of each $S$-wave amplitude matrix should be $\in(- 1/2\,, 1/2)$.
The $S$-wave amplitude matrices are due to fourteen neutral, eight singly-charged, and three doubly-charged scalar channels. 
They read
\beqs
\beqn
\textrm{neutral}~~a_0^0&:& | \pi_i^+ \pi_i^- \rangle\,,\qquad | \pi_1^\pm \pi_2^\mp \rangle \,, \qquad  \frac{1}{\sqrt{2} } | \pi_i^0 \pi_i^0 \rangle\,, \qquad \frac{1}{\sqrt{2} } | h_i h_i \rangle\,, \non
&& | h_i \pi_i^0 \rangle \,, \qquad | \pi_1^0 \pi_2^0 \rangle\,, \qquad | h_1 h_2 \rangle\,,\non
&&  | h_1 \pi_2^0 \rangle\,, \qquad  | h_2 \pi_1^0 \rangle\,,\\
\textrm{singly-charged}~~a_0^+&:& | \pi_i^+ \pi_i^0  \rangle\,, \qquad | \pi_i^+ h_i^0  \rangle\,, \non
&&| \pi_1^+ \pi_2^0  \rangle\,, \qquad | \pi_2^+ \pi_1^0  \rangle\,, \qquad | \pi_1^+ h_2  \rangle\,, \qquad | \pi_2^+ h_1  \rangle \,,\\
\textrm{doubly-charged}~~a_0^{++}&:& \frac{1}{\sqrt{2}} | \pi_1^\pm \pi_1^\pm \rangle \,,\qquad  \frac{1}{\sqrt{2}} | \pi_2^\pm \pi_2^\pm \rangle \,, \qquad  | \pi_1^\pm \pi_2^\pm \rangle \,.
\eeqn
\eeqs
The $S$-wave amplitude matrices for three different channels are expressed as
\beqs
\beqn
&&a_0^0 = \frac{1}{16\pi} {\rm diag}(X_{4\times 4}\,, Y_{4\times 4}\,, Z_{3\times 3}\,, Z_{3\times 3})\,,\\
&& a_0^+ = \frac{1}{16\pi} {\rm diag} (Y_{4\times 4}\,, Z_{3\times 3}\,, \lambda_3 - \lambda_4)\,, \\
&& a_0^{++}=  \frac{1}{16\pi} Z_{3\times 3}
\eeqn
\eeqs
where the expressions for the submatrices of $(X_{4\times 4}\,, Y_{4\times 4}\,, Z_{3\times 3})$ are given in the Ref.~\cite{Kanemura:2015ska}.

\begin{figure}
\centering
\includegraphics[width=6.5cm,height=4.5cm]{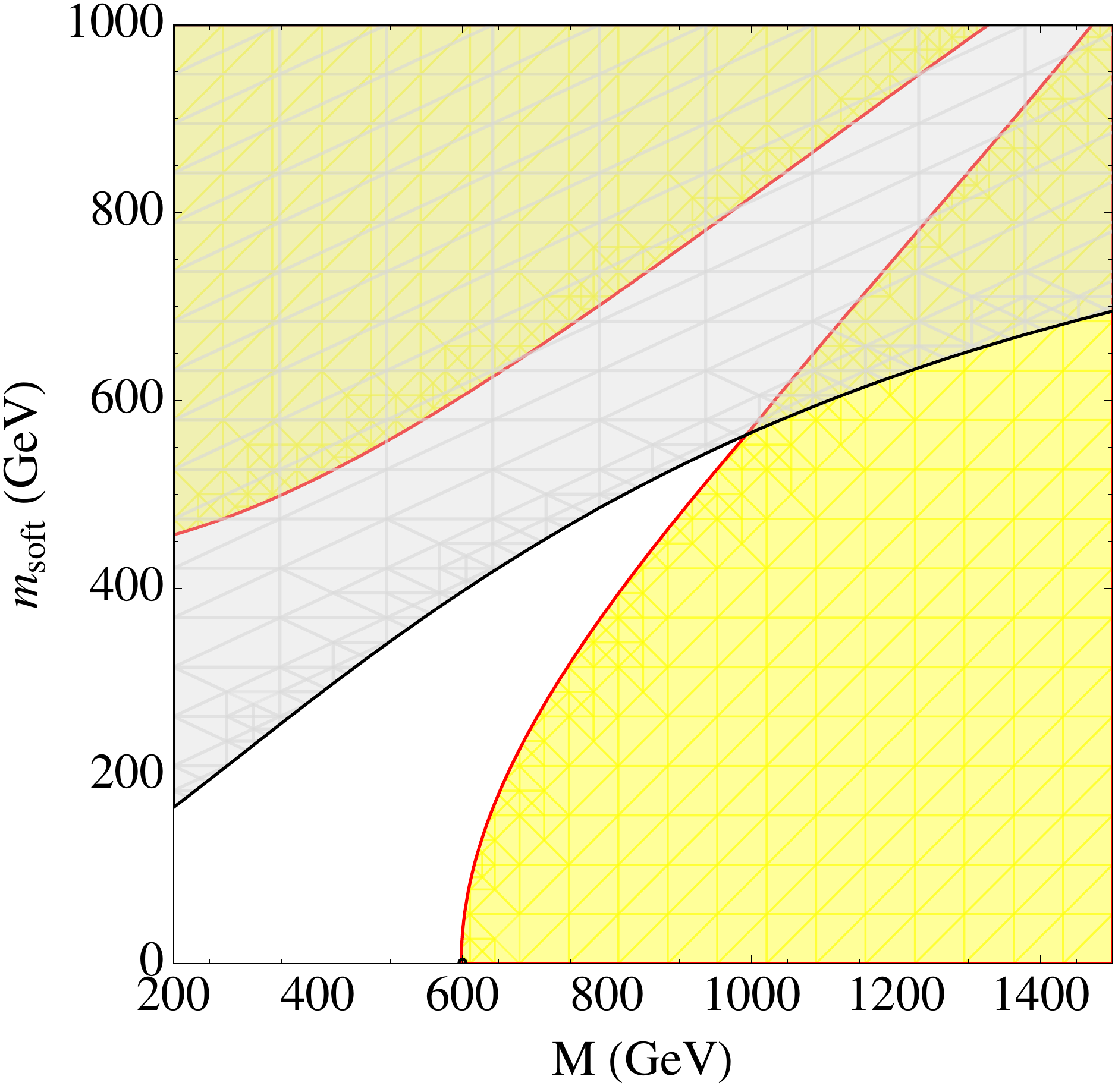}
\includegraphics[width=6.5cm,height=4.5cm]{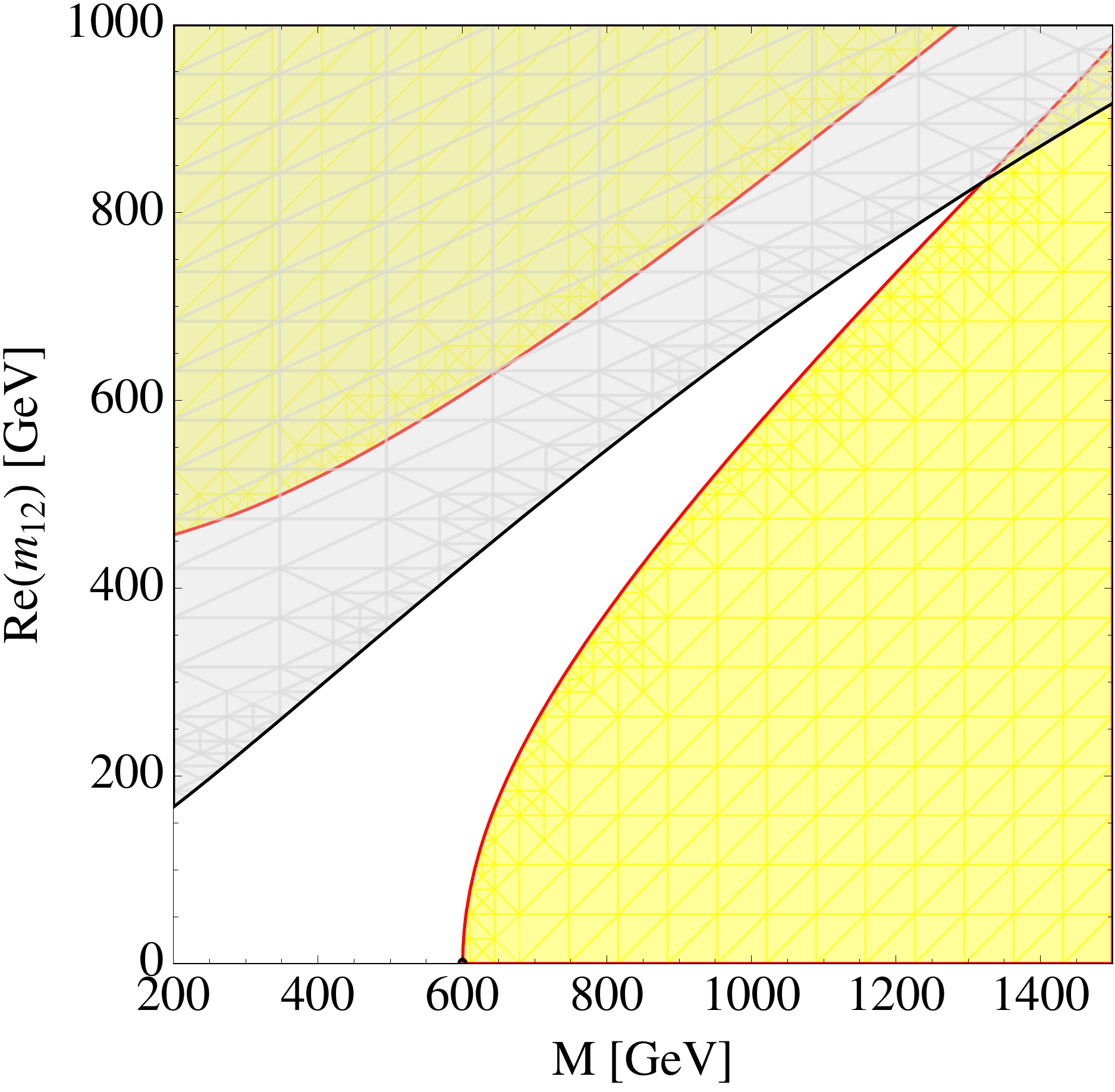}
\caption{\label{fig:MM} 
The combined perturbative unitarity and stability bounds on the $(M\,, m_{\rm soft} )$ plane for the CPV 2HDM-II.
Left: the 2HDM-II with the fixed inputs of $(|\alpha_b|\,,t_\beta)=(0.1\,, 1.0)$, right: the 2HDM-II with the fixed inputs of $( |\alpha_b| \,,t_\beta)=(0.05\,, 1.0)$.
The yellow shaded regions are excluded by the unitarity bounds, and the gray shaded regions are excluded by the stability bounds.
}
\end{figure}

The stability constraints require a positive 2HDM potential for large values of Higgs fields along all field space directions. 
Collectively, they lead to the following conditions
\beqn\label{eq:stability} 
&&\lambda_{1\,,2}>0\,,\qquad \lambda_3 > - \sqrt{\lambda_1 \lambda_2}\,, \qquad \lambda_3 + \lambda_4 - |\lambda_5 | > - \sqrt{ \lambda_1 \lambda_2 }\,,
\eeqn
with $\lambda_{6\,,7}=0$ assumed. 
The combined constraints from the perturbative unitarity and stability to the $(M\,, m_{\rm soft} )$ parameter regions for the CPV 2HDM-II are shown in Fig.~\ref{fig:MM} with the fixed input parameters of $(|\alpha_b|\,,t_\beta)=(0.1\,, 1.0)$ (left panel) and $(|\alpha_b|\,,t_\beta)=(0.05\,, 1.0)$ (right panel).
It turns out that the combined perturbative unitarity and stability put upper bounds to the heavy Higgs boson masses of $M\lesssim 1.0\,\TeV$ for $|\alpha_b| =0.1$, or $M\lesssim 1.2\,\TeV$ for $|\alpha_b|=0.05$.
The stability constraints of \eqref{eq:stability} bound the soft mass term of $m_{\rm soft}$ from above. 
As seen from Eqs.~\eqref{eqs:lambdas}, very large values of $m_{\rm soft}$ will pull $\lambda_{1\,,2}$ into the negative regions, which violate the conditions described by Eqs.~\eqref{eq:stability}.
Later, we will find that the Higgs cubic self couplings, such as $\lambda_{113}$ in our case, become enhanced with the large soft mass inputs of $m_{\rm soft}$ when they are close to the stability boundary.

\subsection{The LHC searches for heavy Higgs bosons}

The constraints to the signal strengths of the $125\,\GeV$ SM-like Higgs boson $h_1$ and the eEDM put bounds to the parameters of $(|\alpha_b|\,, t_\beta)$.
The unitarity and stability constraints put upper bounds to the mass input parameters of $(M\,,  m_{\rm soft})$.
Below, we take into account the constraints from the $7\oplus 8\,\TeV$ LHC searches for the heavy Higgs bosons in the 2HDM spectrum. 
Such constraints were previously given in Ref.~\cite{Chen:2015gaa}, where authors included the constraints from $h_{2\,,3}\to WW/ZZ$ and $h_{2\,,3}\to Z h_1 \to \ell^+ \ell^- b \bar b$ final states. 
Additionally, there have been recent experimental searches to the $hh\to b \bar b \gamma\gamma$ final states from both ATLAS and CMS collaborations, which are included in our studies.

\subsubsection{The heavy Higgs productions}

The cross sections of the heavy Higgs bosons via the ggF channel can be rescaled from the SM-like Higgs production with the same mass as
\beqn\label{eq:ggtohi}
\frac{\sigma[gg\to h_i ]}{\sigma[gg\to h_{\rm SM}]}&=& \frac{ \Big| c_{t\,,i} A_{1/2}^H(\tau_t^i ) + c_{b\,,i} A_{1/2}^H(\tau_b^i )   \Big|^2 +  \Big| \tilde c_{t\,,i} A_{1/2}^A(\tau_t^i ) +  \tilde c_{b\,,i}  A_{1/2}^A(\tau_b^i )   \Big|^2  }{ \Big|  A _{1/2}^H(\tau_t^i) + A_{1/2}^H (\tau_b^i )  \Big|^2 }\,,
\eeqn
with the variable of
\beqn
&&\tau_f^i\equiv \frac{M_i^2 }{ 4 m_f^2}\,,\qquad f=t\,,b\,.
\eeqn
The cross sections of the heavy Higgs bosons via the VBF channel can be rescaled from the SM-like Higgs production with the same mass as
\beqn
\frac{\sigma[qq \to qq h_i ]}{\sigma[qq \to qq h_{\rm SM}]}&=& a_i^2\,.
\eeqn

\subsubsection{The heavy Higgs decays}

Here, we list the partial decay widths of the heavy neutral Higgs bosons at the leading order (LO).
The partial decay widths into the gauge bosons are
\beqn\label{eq:hitoVV_wid}
\frac{\Gamma[h_i \to VV]}{\Gamma[h_{\rm SM}\to VV] }&=& a_i^2\,,
\eeqn
with $V=(W^\pm \,,Z)$.
The partial decay widths into the SM fermions are 
\beqn\label{eq:hitoff_wid}
\frac{\Gamma[h_i \to f \bar f ] }{ \Gamma[h_{\rm SM}\to f \bar f] }&=& ( c_{f\,,i} )^2 + ( \tilde c_{f\,,i}  )^2.
\eeqn
We also consider the non-standard decay modes of the heavy Higgs bosons, which include $h_i \to h_1 Z$, $H^\pm \to h_1 W^\pm$, and $h_i\to h_1 h_1$. 
Their partial decay widths are
\beqs
\beqn
\Gamma[h_i \to h_1 Z]&=& \frac{  | g_{i1z} |^2 }{16\pi M_i } \sqrt{ \Big( 1 - \frac{(M_1 + m_Z)^2}{M_i^2}  \Big)  \Big(  1 - \frac{(M_1 - m_Z)^2}{M_i^2}  \Big)  } \non
&&\times \Big[  \frac{1}{m_Z^2} ( M_i^2  - M_1^2 )^2 - (2 M_i^2 + 2 M_1^2 - m_Z^2)  \Big]\,,\label{eq:hitoh1Z_wid}\\
\Gamma[h_i \to h_1 h_1]&=& \frac{\lambda_{11i}^2 }{ 4\pi M_i } \sqrt{1 - \frac{4 M_1^2 }{M_i^2}  }\,,\label{eq:hitoh1h1_wid}
\eeqn
\eeqs
where $g_{iz1}=(e/s_{2W} ) [ ( - s_\beta \mR_{11} + c_\beta \mR_{12} ) \mR_{i3} - ( - s_\beta \mR_{i1} + c_\beta \mR_{i2} ) \mR_{13}  ] $.
The cubic self couplings of $\lambda_{11i}$ are obtained in Eq.~\eqref{eq:lambda_ijk} from the Lagrangian terms in Eq.~\eqref{eq:L3s}, and their expansions in terms of the CPV mixing angle $\alpha_b$ are given in Eqs.~\eqref{eqs:lambda_ijk_expansion} later.
By fixing the parameter choices of the alignment limit and $\alpha_c=0$, we find the non-vanishing couplings of $g_{2z1}= - (e/s_{2W} ) s_{\alpha_b}$ and $\lambda_{113}\neq 0$.

\subsubsection{The experimental search bounds}

\begin{figure}
\centering
\includegraphics[width=6.5cm,height=4.5cm]{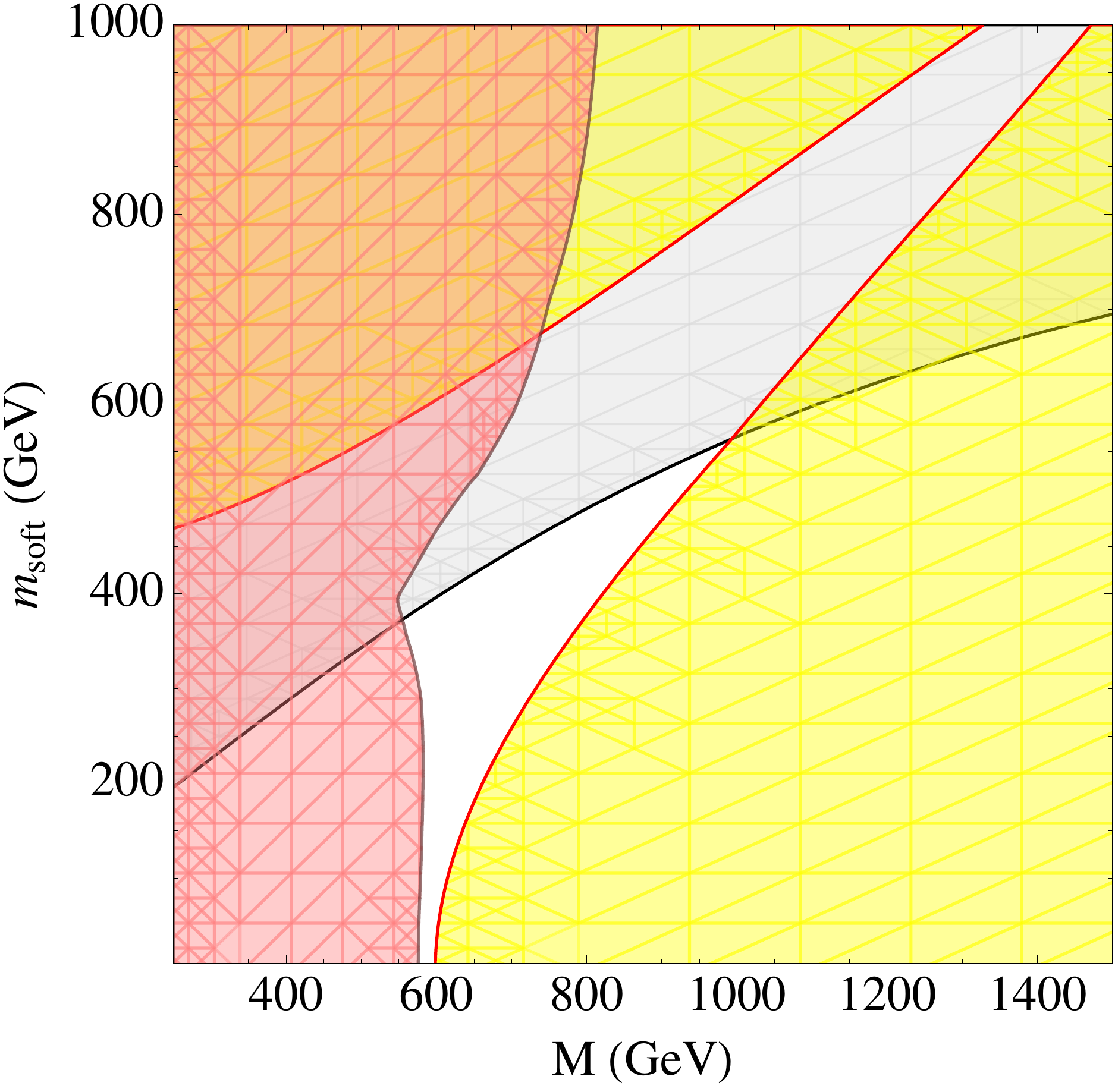}
\includegraphics[width=6.5cm,height=4.5cm]{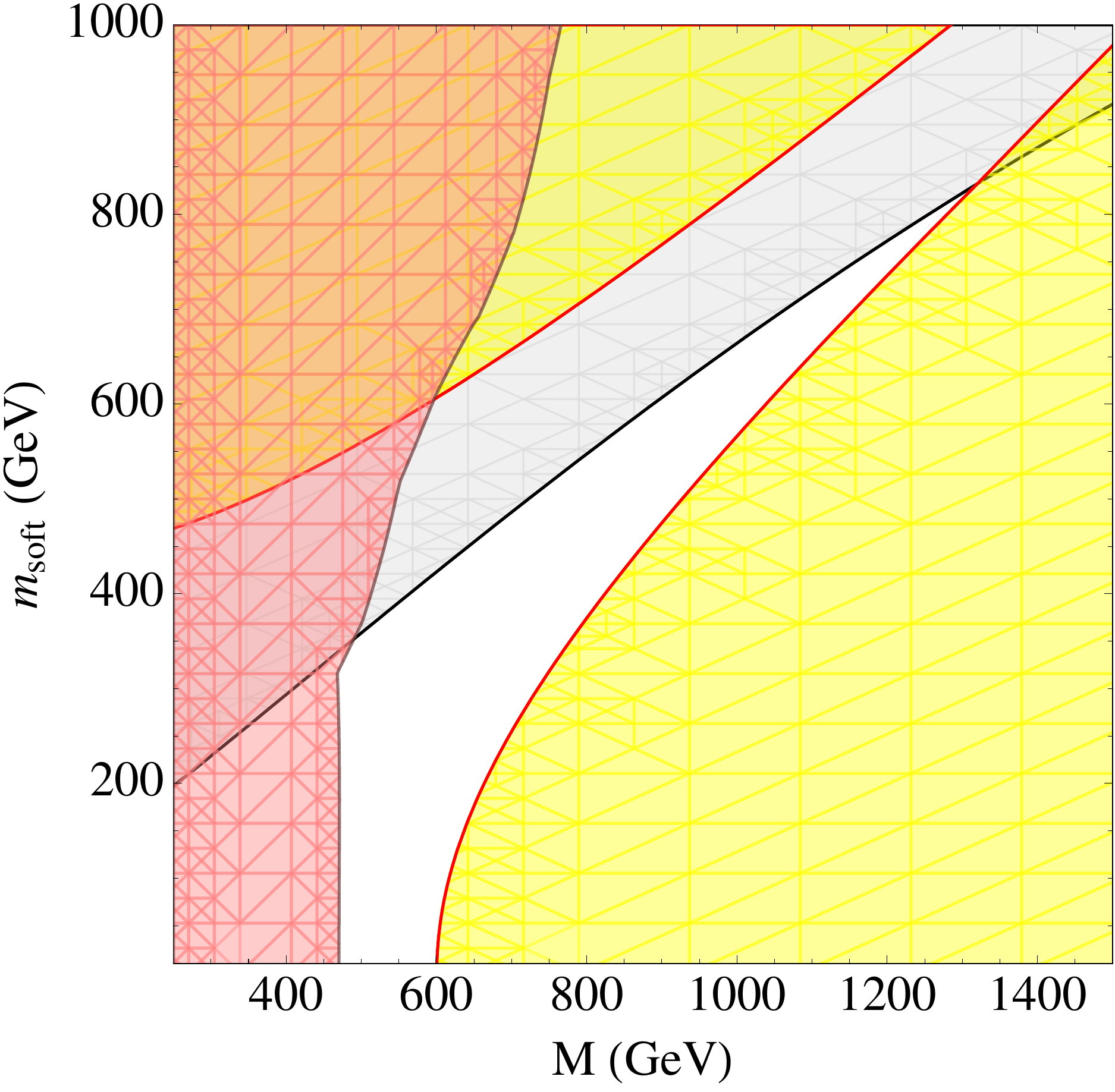}
\caption{\label{fig:h2h3Exclusion} 
The combined unitarity and stability bounds on the $(M\,, m_{\rm soft} )$ plane for the CPV 2HDM-II, with fixed parameter of $t_\beta=1.0$.
Left: $|\alpha_b|=0.1$, right: $|\alpha_b| = 0.05$.
The pink shaded regions are excluded by the LHC searches for the heavy Higgs bosons.
}
\end{figure}

The current LHC experimental searches for the heavy Higgs bosons are performed via the $(WW\,, ZZ)$ final states~\cite{Khachatryan:2015cwa,Aad:2015kna}, the $H\to hh\to b \bar b + \gamma\gamma$~\cite{Aad:2014yja, Khachatryan:2016sey}, and $A\to hZ\to (b \bar b + \ell^+ \ell^-/\tau^+ \tau^- +\ell^+ \ell^-)$~\cite{Aad:2015wra,Khachatryan:2015tha}. 
Since we always assume that $M_2=M_3$, the constraints to the heavy Higgs boson searches at the LHC are imposed to the cross sections of $\sigma[pp\to h_2/h_3\to XX ]$
\beqn
\sigma[pp\to h_2/h_3\to XX ]&=&\sigma [gg\to h_2]\times {\rm Br}[h_2 \to XX]\non
& +& \sigma [gg\to h_3]\times {\rm Br}[h_3 \to XX]\,,
\eeqn
where we consider the leading production channel of ggF obtained from Eq.~\eqref{eq:ggtohi}. 
The decay branching ratios are obtained from the partial decay widths of Eqs.~\eqref{eq:hitoVV_wid}, \eqref{eq:hitoff_wid}, \eqref{eq:hitoh1Z_wid}, and \eqref{eq:hitoh1h1_wid} evaluated at the LO.
We find the most stringent constraint to the heavy Higgs boson searches are from the recent CMS searches for the resonances with two SM-like Higgs bosons in Ref.~\cite{Khachatryan:2016sey}.
By converting all heavy Higgs boson constraints to the $(M\,, m_{\rm soft})$ plane, we find the mass regions of $M_{2\,,3}\lesssim 600\,\GeV$ are excluded for $|\alpha_b|=0.1$, or $M_{2\,,3}\lesssim 500\,\GeV$ are excluded for $|\alpha_b|=0.05$, respectively.
The current $B$-physics data also excludes the charged Higgs boson mass greater than $M_\pm \sim 340\,\GeV$ for 2HDM-II~\cite{WahabElKaffas:2007xd,Mahmoudi:2009zx}.
Combining with the previous unitarity and stability constraints, we display the allowed parameter regions of $(M\,, m_{\rm soft} )$ in Fig.~\ref{fig:h2h3Exclusion}.
Accordingly, we consider two scenarios of
\beqn\label{eq:MassRange}
{\rm (i)}&:& |\alpha_b| = 0.1\,,~{\rm  with} ~~M_{2\,,3}\in (600\,\GeV \,,1000\,\GeV)\,,\non
{\rm (ii)}&:& |\alpha_b| =0.05\,, ~{\rm with} ~~M_{2\,,3}\in (500\,\GeV \,,1200\,\GeV)\,,
\eeqn
for the Higgs pair productions at the future high-energy $e^+ e^-$ and $pp$ colliders. 
A set of benchmark models for the $|\alpha_b|=0.1$ and $|\alpha_b|=0.05$ cases are listed in Table.~\ref{tab:benchmark}, where the soft mass terms of $m_{\rm soft}$ are chosen to be close to the stability boundary for each heavy Higgs boson mass.
In the next section, we will study the Higgs pair productions at the future $e^+ e^-$ and $pp$ collider experiments based on these benchmark models.

\begin{table}[htb]
\begin{center}
\begin{tabular}{c|c|c|c|c|c|c}
\hline
\hline
  &  \multicolumn{3}{|c|}{$ |\alpha_b|=0.1$} & \multicolumn{3}{|c}{$|\alpha_b|=0.05$}  \\\hline
 $M_2=M_3({\rm GeV})$   &  $m_{\rm soft}({\rm GeV})$ &  $\lambda_{111}({\rm GeV})$  & $\lambda_{113}({\rm GeV})$   &  $m_{\rm soft}({\rm GeV})$  & $\lambda_{111}({\rm GeV})$   & $\lambda_{113}({\rm GeV})$  \\ \hline
 $500$ & $...$ & $...$ & $...$  &  $350$  & $29.37$  & $-70.33$  \\
 $600$  & $400$ & $19.45$ & $-173.75$ & $420$  & $28.28$ & $-102.66$ \\
 $700$  & $440$ & $16.80$ & $-200.04$  & $480$   & $27.19$ & $-133.01$ \\
 $800$  & $480$  & $13.89$ &  $-227.22$ & $540$  & $25.96$ &  $-167.14$ \\
 $900$  & $520$  & $10.74$ & $-255.31$  & $600$  & $24.57$ & $-205.05$ \\
 $1000$ & $560$  & $7.33$ & $-284.30$  & $660$  & $23.05$ & $-246.72$  \\
 $1100$ & $...$  & $...$ &  $...$ & $710$  & $21.66$ & $-280.60$  \\ 
 $1200$ & $...$  & $...$ &  $...$ & $770$  & $19.87$ & $-328.86$  \\   
 \hline\hline
\end{tabular}
\caption{
The CPV 2HDM-II benchmark models for the CPV mixing angles of $|\alpha_b|=0.1$ and $|\alpha_b|=0.05$.
The heavy Higgs boson mass ranges are chosen according to Eq.~\eqref{eq:MassRange}. 
The non-vanishing Higgs cubic self couplings of $\lambda_{111}$ and $\lambda_{113}$ are listed for each model. 
}
\label{tab:benchmark}
\end{center}
\end{table}

\subsection{The EW precision constraints}

The Peskin-Takeuchi parameters of $(S\,,T)$ for the EW precision tests were obtained in Refs.~\cite{Grimus:2008nb,Branco:2011iw,He:2001tp,Haber:2010bw,Hernandez:2015rfa,Hernandez:2015dga} for the 2HDM.
In our simplified case with the alignment limit, the degenerate masses of $M_2=M_3=M_\pm$, and $\alpha_c=0$, they read 
\beqs\label{eqs:ST}
\beqn
\Delta S&=&\frac{1}{96\,\pi^2 c_W} \frac{m_W^2}{v^2} \Big\{  c_{2w}^2 G(M_\pm^2\,, M_\pm^2\,, m_Z^2) + \Big[ G(M_1^2\,, M_2^2\,, m_Z^2) + \hat G( M_3^2\,, m_Z^2)  \Big] s_{\alpha_b}^2 \non
&+& \Big[  \hat G(M_1^2\, m_Z^2 ) + G(M_2^2\,, M_3^2\,, m_Z^2 ) \Big] c_{\alpha_b}^2 + \log\Big(  \frac{M_1^2 M_2^2  M_3^2	}{M_\pm^6} \Big) \non
&-& \Big[  \hat G(M_{H\,, {\rm ref}}^2\,, m_Z^2 ) + \log \Big( \frac{M_{H\,, {\rm ref}}^2}{ M_\pm^2 }  \Big)  \Big] \Big\} \,,\\
\alpha \Delta T&=& \frac{1}{6\pi^2\, v^2} \Delta_1 (m_W - m_Z) s_{\alpha_b}^2\,.
\eeqn
\eeqs
for a reference value of the SM Higgs boson mass $M_{H\,,{\rm ref}}=125\,\GeV$.
Here, we denote $\Delta_1\equiv M_\pm - M_1 $, and the functions of $G(x,y,z), \hat{G}(x,y)$ are given in~\cite{Grimus:2008nb}.
By employing the current Gfitter fit to the EW data \cite{Baak:2014ora}, the parameters are founded to be constrained by $T$ parameter mostly for the CPV parameter $\alpha_b$ allowed by Fig.~\ref{fig:h125sigEDM}, and the degenerate masses of heavy Higgs bosons relax the constraints again.


\section{Higgs Pair Productions at The Colliders}
\label{section:collider}

In this section, we study the SM-like Higgs pair productions in the framework of the CPV 2HDM. 
The SM-like Higgs cubic self coupling of $\lambda_{111}$ are modified due to the varying inputs of the soft mass term and the CPV mixing angle. 
Therefore, we will discuss the precision measurement of $\lambda_{111}$ at the future $e^+ e^-$ colliders for the benchmark models in Table.~\ref{tab:benchmark}.
We will also focus on the most dominant channel for the resonance contributions, namely the ggF process at the hadron colliders, which include the LHC 14 TeV and the future SppC/Fcc-hh 100 TeV runs.

\subsection{The Higgs cubic self couplings}

Before evaluating the cross sections of the Higgs pair productions, it is necessary to look at the behaviors of the relevant Higgs cubic self couplings of $\lambda_{11i}\,(i=1,2,3)$. 
Following the previous constraints, we fix the parameters of $(\alpha\,,t_\beta)=(-\pi/4\,,1.0)$, and keep the input of $\alpha_c=0$.
With these assumptions, we find that only $\lambda_{111}$ and $\lambda_{113}$ survive, and $\lambda_{112}$ is always vanishing.
Their explicit expressions in the mass eigenbasis read
\beqs\label{eqs:lambda_ijk}
\beqn
\lambda_{111}&=& \frac{c_{\alpha_b}}{2v}\Big[  M_1^2\,( s_{\alpha_b}^4 + c_{\alpha_b}^4 + s_{2\alpha_b}^2) + 2 M_3^2\,s_{\alpha_b}^4  - 4 m_{\rm soft}^2\, s_{\alpha_b}^2 \Big]\,,\\
\lambda_{113}&=& \frac{s_{\alpha_b}}{8v}  \Big[ M_1^2\, ( 3\, c_{4 \alpha_b} + 8\, c_{2\alpha_b} -3 ) - M_3^2 \,( 3\, c_{4 \alpha_b} - 4\, c_{2\alpha_b} -3 )\non
&-& 8 m_{\rm soft}^2\, ( 3\, c_{2\alpha_b} +1)   \Big] \,.
\eeqn
\eeqs
Since the CPV mixing angle of $\alpha_b$ is typically small by imposing the eEDM constraints, it is also useful to expand the cubic couplings in terms of the $\alpha_b$ angle as follows
\beqs\label{eqs:lambda_ijk_expansion}
\beqn
\lambda_{111}&\simeq& \frac{M_1^2}{2v } + \frac{3 M_1^2 - 8\, m_{\rm soft}^2 }{4v}\cdot \alpha_b^2 + \mO(\alpha_b^4)\,,\label{eq:lambda111}\\
\lambda_{113}&\simeq&\frac{2 M_1^2 + M_3^2 - 8\, m_{\rm soft}^2 }{2v}\cdot \alpha_b + \mO(\alpha_b^3)\,.\label{eq:lambda113}
\eeqn
\eeqs
The Higgs cubic self coupling of $\lambda_{111}$ starts with the SM predicted values of  $\lambda_{hhh}^{\rm SM}\simeq 32\,\GeV$, plus the higher order corrections of $\mO(\alpha_b^2)$.
The overall magnitude of $\lambda_{113}$ is controlled by the size of the CPV mixing angle $\alpha_b$.
Hence, one can expect that the improvement in the precisions of the eEDM measurements will reduce the size of the heavy resonance contributions to the Higgs pair productions via the ggF process.

\begin{figure}
\centering
\includegraphics[width=6.5cm,height=4.5cm]{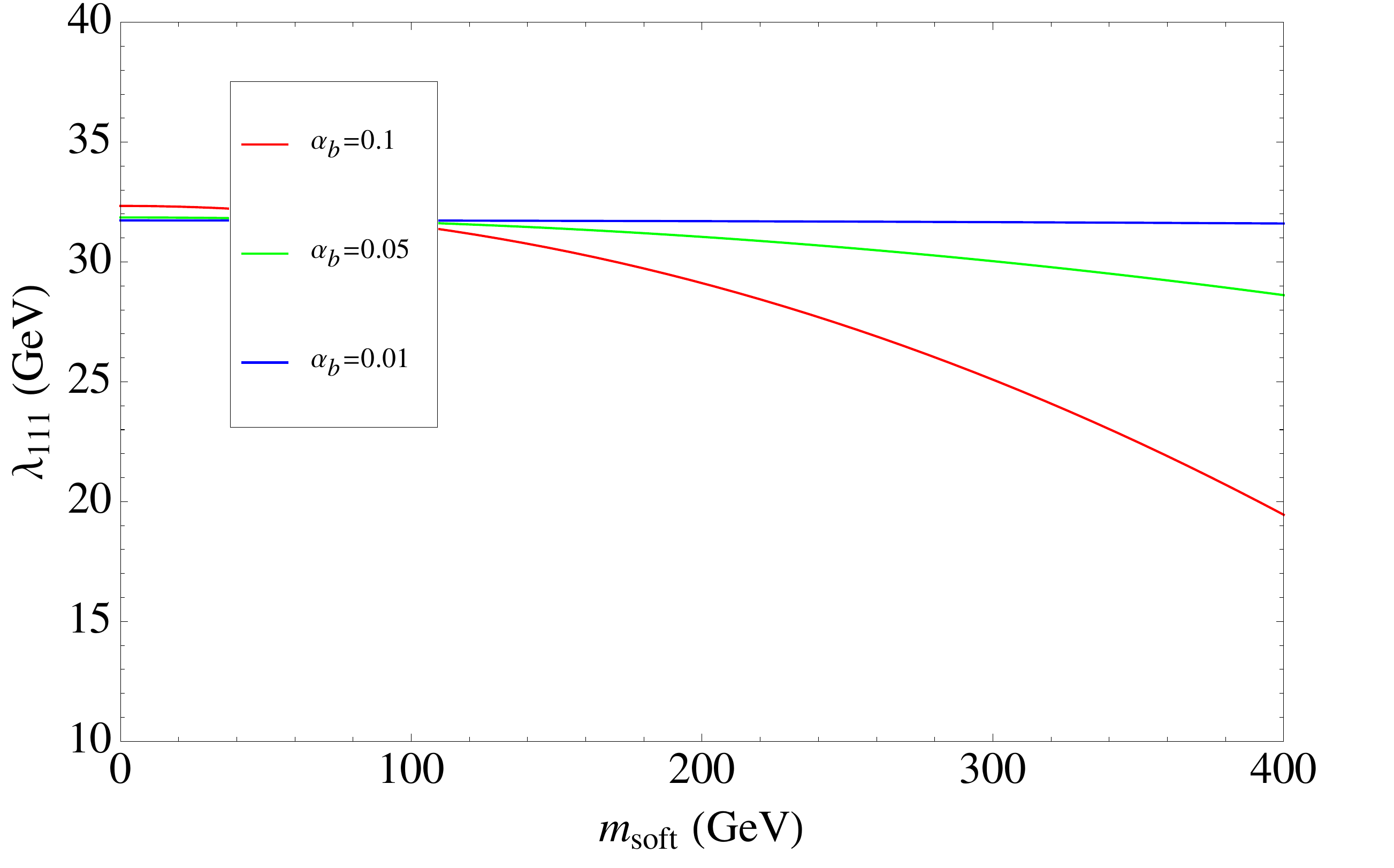}
\includegraphics[width=6.5cm,height=4.5cm]{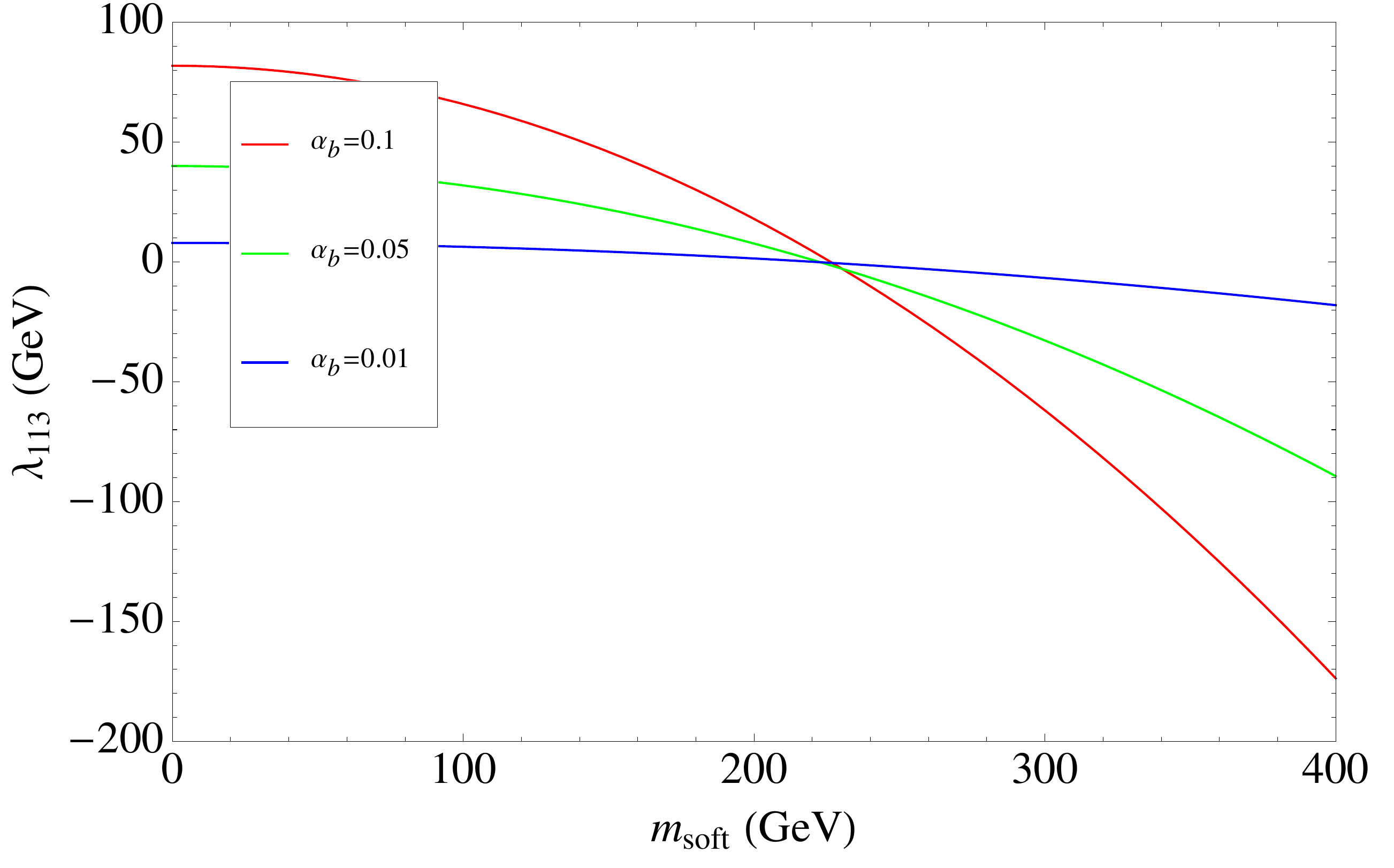}
\caption{\label{fig:lambdas} 
The Higgs cubic self couplings $\lambda_{111}$ (left) and $\lambda_{113}$ (right) versus $m_{\rm soft}$ for the $M_2=M_3=600\,\GeV$ case in the CPV 2HDM-II, with fixed inputs of $\alpha=-\pi/4$ and $t_\beta = 1.0$.
}
\end{figure}

In Fig.~\ref{fig:lambdas}, we plot the Higgs cubic self couplings of $\lambda_{111}$ and $\lambda_{113}$ for the $M_2=M_3=600\,\GeV$ case with different CPV mixing angles of $\alpha_b$ in the CPV 2HDM-II.
The lower and upper bounds of the soft mass inputs $m_{\rm soft}$ in these plots are from the perturbative unitarity and the stability constraints, respectively.
For a fixed input of $\alpha_b$, the Higgs cubic self coupling of $\lambda_{111}$ becomes smaller than the SM predicted value with the increasing inputs of $m_{\rm soft}$.
On the other hand, when the CPV mixing angle becomes as small as $\alpha_b =0.01$, the Higgs cubic self coupling of $\lambda_{111}$ is basically the same as $\lambda_{hhh}^{\rm SM}\simeq 32\,\GeV$.
The other Higgs cubic self coupling of $\lambda_{113}$ also decreases from positive regions to negative regions with the increasing inputs of $m_{\rm soft}$.
Its variation is also controlled by the size of the CPV mixing angle of $\alpha_b$, as seen from its behaviors with the different inputs of the CPV mixing angle of $\alpha_b=(0.1\,, 0.05\,, 0.01)$.
For the $M_2=M_3=600\,\GeV$ case, $\lambda_{113}$ tends to zero when the soft mass term is $m_{\rm soft} \simeq  220\,\GeV$, as can be evaluated from Eq.~\eqref{eq:lambda113}.
Thus, one would expect the corresponding resonance contributions to vanish.
When the soft mass deviates from this value of $m_{\rm soft} \simeq 220\,\GeV$, either increases to the stability boundary or decreases to zero, $|\lambda_{113}|$ increases. 
Correspondingly, one can expect large resonance contributions for such parameter inputs.

\subsection{The precise measurement of $\lambda_{111}$ at the future $e^+ e^-$ colliders}

\begin{figure}
\centering
\includegraphics[width=6.5cm,height=4.5cm]{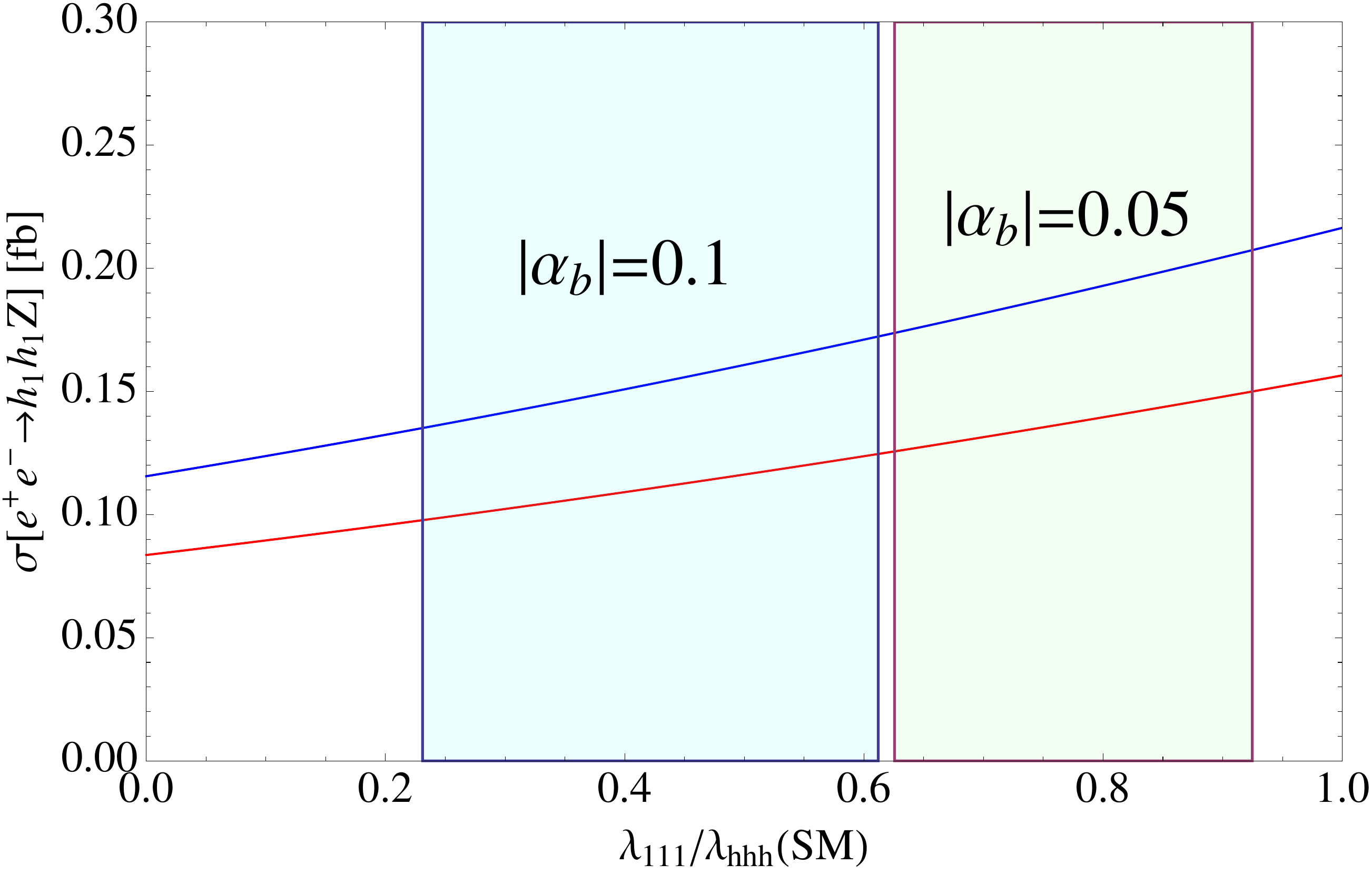}
\includegraphics[width=6.5cm,height=4.5cm]{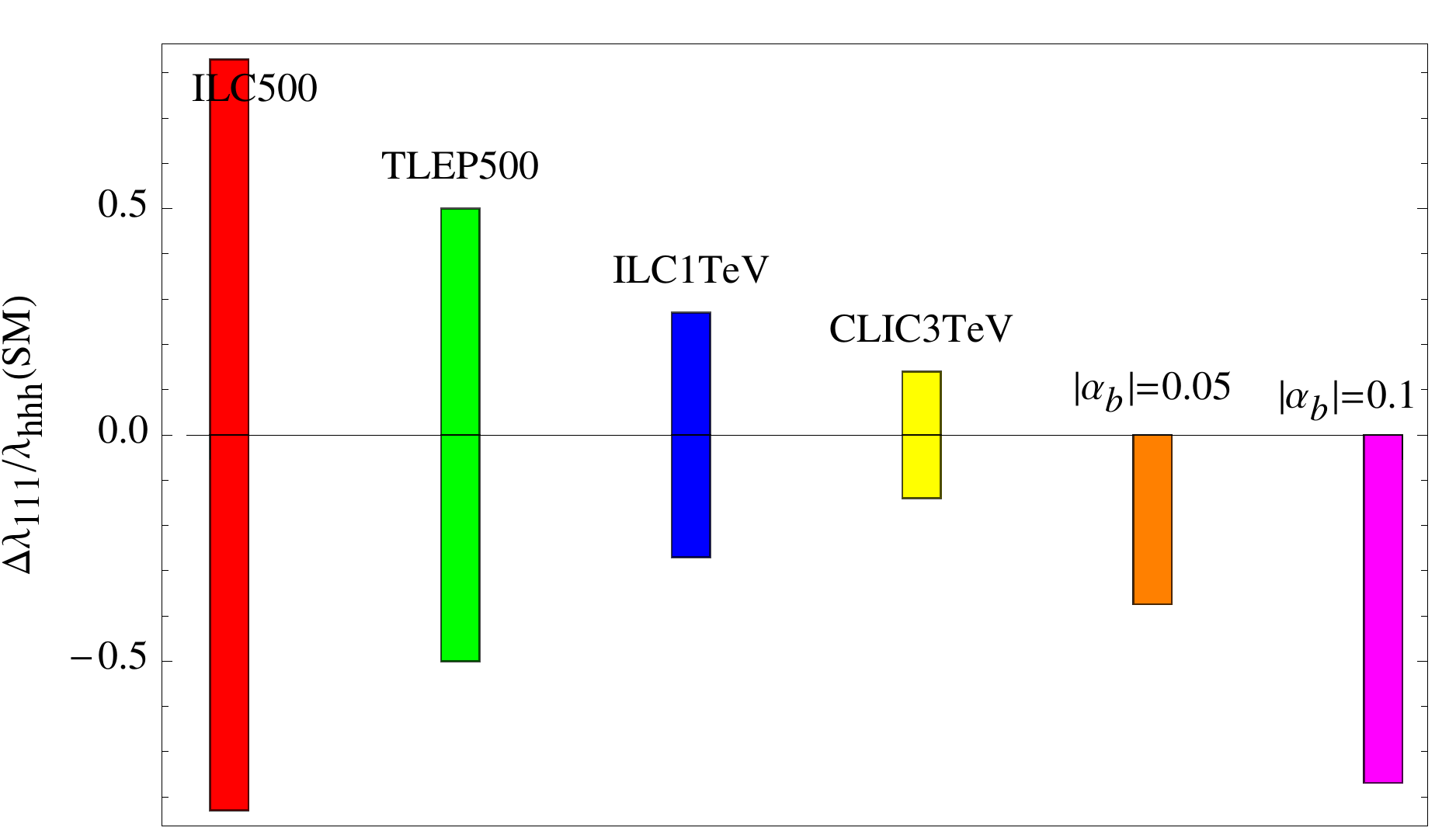}
\caption{\label{fig:eehhZ} 
Left: the cross sections of $\sigma[e^+ e^- \to h_1 h_1 Z]$ at the TLEP (red) and ILC (blue) $500\,\GeV$ versus the different Higgs cubic self couplings.
Right: the expected accuracies on the Higgs cubic self couplings at the future $e^+ e^-$ colliders, and the $\Delta \lambda_{111}/\lambda_{hhh}^{\rm SM}$ for the benchmark models of $|\alpha_b|=0.1$ and $|\alpha_b|=0.05$.
}
\end{figure}

The future high-energy $e^+ e^-$ colliders provide opportunities of measuring the SM-like Higgs cubic self couplings. 
The direct measurements can be achieved via the $e^+ e^- \to h h Z$ process with the center-of-mass energy of $\sqrt{s} = 500\,\GeV$, or via the vector boson fusion process of $e^+ e^- \to hh \nu_e \bar \nu_e$ with the center-of-mass energy of $\sqrt{s}=1\,\TeV$~\cite{Baer:2013cma,Gomez-Ceballos:2013zzn,Tian:2010np}.
The first advantage of the $e^+ e^-$ colliders is that the relevant Higgs-gauge couplings for these processes can be precisely measured to the percentage level at the $\sqrt{s}=240-250\,\GeV$ runs~\cite{Baer:2013cma,Gomez-Ceballos:2013zzn,CEPC-SPPC-pre}.
For the CPV 2HDM, one has the Higgs-gauge couplings of 
\beqn
&& g_{1ZZ} = g_{hZZ}^{\rm SM}\,c_{\alpha_b} = \frac{m_Z^2}{v}\,c_{\alpha_b}\,,\qquad g_{11ZZ}= g_{hhZZ}^{\rm SM} = \frac{m_Z^2}{2v^2}\,.
\eeqn
with $\delta g_{1ZZ}=\Big| g_{1ZZ} - g_{hZZ}^{\rm SM} \Big| < \mO(1\,\%)$ after imposing the eEDM constraints.
The second advantage of the $e^+ e^-$ colliders is that the contributions to the total cross section from the heavy resonance of $h_3$ are typically less than $\mO(10^{-4})$, hence they are negligible.
Therefore, it is a good approximation to assume the SM predicted values for the Higgs-gauge couplings, and only vary the Higgs cubic self coupling of $\lambda_{111}$.
The ratio of the total cross section of $\sigma[e^+ e^- \to h h Z]$ to its SM counterpart can be parametrized as follows
\beqn
\frac{\sigma[e^+ e^- \to h_1 h_1 Z]}{~~~\sigma[e^+ e^- \to h h Z]_{\rm SM}} &=& 0.097\,\xi_{111}^2 + 0.369\,\xi_{111} + 0.534\,,
\eeqn
at the TLEP and ILC $500\,\GeV$ runs, with $\xi_{111} \equiv \lambda_{111}/\lambda_{hhh}^{\rm SM}$.
The total cross sections at the TLEP and ILC $500\,\GeV$ runs versus the ratios of different Higgs cubic self couplings $\lambda_{111}/\lambda_{hhh}^{\rm SM}$ are displayed on the left panel of Fig.~\ref{fig:eehhZ}.
The ranges of $\lambda_{111}$ in two set of benchmark models with $|\alpha_b|=0.1$ and $|\alpha_b|=0.05$ are also shown in the light-blue and light-green shaded regions, respectively.
From the results given in Table.~\ref{tab:benchmark} for the benchmark models, the Higgs cubic self couplings of $\lambda_{111}$ are always smaller than the SM predicted values.
Thus, the corresponding cross sections of $\sigma[e^+ e^- \to h_1 h_1 Z]$ are smaller than the SM predictions at the TLEP and the ILC.
On the right panel of Fig.~\ref{fig:eehhZ}, we display the expected accuracies on the Higgs cubic self couplings for ILC500 (with $\int \mL dt=0.5\,\ab^{-1}$), TLEP500 (with $\int \mL dt=1\,\ab^{-1}$), ILC $1\,\TeV$ (with $\int \mL dt=1\,\ab^{-1}$), and CLIC $3\,\TeV$ (with $\int \mL dt=2\,\ab^{-1}$).
The deviations of the Higgs cubic self couplings $\Delta \lambda_{111}/\lambda_{hhh}^{\rm SM}$ corresponding to the benchmark models of $|\alpha_b|=0.1$ and $|\alpha_b|=0.05$ are shown for comparison.
For the $|\alpha_b|=0.1$ case, the largest deviations of $\lambda_{111}$ can be probed with the accuracies reached by the TLEP $500\,\GeV$; while for the smaller CPV mixing angle of $|\alpha_b|=0.05$ case, the largest deviations of $\lambda_{111}$ can be probed with the accuracies reached by the ILC $1\,\TeV$.

\subsection{The $p p\to h_1 h_1$ in the CPV 2HDM}

The parton-level differential cross sections of the Higgs pair production for both SM Higgs and BSM Higgs bosons via the ggF process were previously derived in Refs.~\cite{Glover:1987nx, Plehn:1996wb, Dawson:1998py, Djouadi:1999rca}. 
For the productions of the SM-like Higgs boson pairs, its differential cross section reads
\beqn\label{eq:dsigmahh_SM}
\frac{d \hat \sigma}{ d \hat t} [ g g \to h h ]&=& \frac{ G_F^2 \alpha_s^2 }{ 512 (2\pi)^3} \sum_{q} \Big[  \Big|   ( C_\triangle^h  F_\triangle^h  +  C_\Box^{hh} F_\Box^{hh} ) \Big|^2 +  \Big|  C_\Box^{hh} G_\Box^{hh}  \Big|^2  \Big]\,,
\eeqn
where the dominant contributions are due to the top-quark loops.
The form factors of $(F_\triangle\,, F_\Box\,, G_\Box)$ are from the loop integrals of the triangle diagrams, the $J=0$ partial wave of the box diagrams, and the $J=2$ partial wave of the box diagrams.
Their explicit expressions are summarized in the appendix of Ref.~\cite{Plehn:1996wb}.
The relevant coefficients are given by
\beqs
\beqn
&& C_\triangle^h = \frac{ (6\lambda_{hhh}v )\, \xi_h^q   }{ \hat s - M_h^2 + i M_h \Gamma_h } \,,\\
&&C_\Box^{hh} =(\xi_h^q )^2 \,,
\eeqn
\eeqs
with $\lambda_{hhh}$ and $\xi_h^q$ representing the Higgs cubic self couplings and the dimensionless Yukawa couplings of the SM-like Higgs boson, respectively.
For the SM case, these couplings are
\beqn
&& \lambda_{hhh}^{\rm (SM) } = \frac{ M_h^2 }{2 v}\,, \qquad ( \xi_h^q )^{\rm (SM)} = 1\,.
\eeqn
The LO total cross sections for the LHC 14 TeV runs and the SppC 100 TeV runs can be estimated by using Madgraph 5~\cite{Alwall:2014hca} as follows
\beqn\label{eq:sigmahh_SM}
&&\sigma_{\rm LO}^{14}[pp\to hh]= 17.34\,\fb\,,\qquad \sigma_{\rm LO}^{100}[pp\to hh]= 806.6\,\fb\,.
\eeqn

For the most general case in the CPV 2HDM, all neutral Higgs bosons of $h_i$ have both CP-even and CP-odd Yukawa couplings. 
Furthermore, the heavy resonances enter into the Higgs pair productions. 
The corresponding differential cross sections at the parton level can be generalized from the results in the appendix of Ref.~\cite{Plehn:1996wb} for the different CP combinations of the final-state $h_1\, h_1$, which are expressed as follows
\beqn\label{eq:dsigmahh_CPV}
\frac{d\hat \sigma}{d \hat t}[g g\to h_1 h_1]&=& \frac{G_F^2 \alpha_s^2 }{512\, (2\pi)^3} \sum_{q} \Big[  \Big|  (\sum_{h_i=h_1\,,h_3} C_\triangle^{h_i}  )  F_\triangle^{h}  + C_\Box^{hh} F_\Box^{hh} + C_\Box^{AA} F_\Box^{AA}  \Big|^2 \non
&+&  \Big| ( \sum_{h_i=h_1\,,h_3} \tilde C_\triangle^{h_i} ) F_\triangle^A  + C_\Box^{hA} F_\Box^{hA} \Big|^2\non
& +& \Big| C_\Box^{hh} G_\Box^{hh} + C_\Box^{AA} G_\Box^{AA}  \Big|^2  + \Big|  C_\Box^{hA} G_\Box^{hA}  \Big|^2   \Big]\,.
\eeqn
The relevant couplings are
\beqs\label{eqs:h1h1_couplings}
\beqn
&& C_\Box^{hh}= (c_{q\,,1})^2\,,\\
&& C_\Box^{AA} = (\tilde c_{q\,,1} )^2\,,\\
&& C_\Box^{hA}= c_{q\,,1}  \tilde c_{q\,,1}\,,\\
&& C_\triangle^{h_i} = \frac{  (g_{11i} v) c_{q\,,i}}{\hat s - M_{i}^2 + i M_{i} \Gamma_{i}  }  \,,\\
&& \tilde C_\triangle^{h_i} = \frac{  (g_{11i} v)  \tilde c_{q\,,i} }{\hat s - M_{i}^2 + i M_{i} \Gamma_{i}  } \,,
\eeqn
\eeqs
with $g_{111}=6\lambda_{111}$ and $g_{113}=4\lambda_{113}$.
To evaluate the cross sections, we implement all couplings given in Eqs.~\eqref{eqs:h1h1_couplings} into the FeynRules~\cite{Christensen:2008py}, and pass the UFO model files into the Madgraph 5.

Now we present the results of the Higgs pair productions in the CPV 2HDM, by combining all previous constraints.
As one can learn from Eq.~\eqref{eq:dsigmahh_CPV}, the cross sections of $\sigma[pp\to h_1 h_1]$ get modified from their SM counterparts due to: (i) the modification of the Higgs cubic self coupling $\lambda_{111}$, (ii) the modifications of the top quark Yukawa couplings, and (iii) the additional resonance contributions.
Through the signal fit to the $125\,\GeV$ SM-like Higgs boson $h_1$ and the eEDM constraints, the dimensionless Higgs Yukawa couplings are bounded such that $\delta c_{f\,,1} < 1\,\%$ and $\tilde c_{f\,,1}\sim - 0.1$.
Therefore, the box diagram contributions are envisioned to approach to the SM predicted values.
From the previous estimation of the Higgs cubic self couplings for the $M_2=M_3=600\,\GeV$ case, we may either have the large resonance contributions or go to the regions with the vanishing resonance contributions of $(\lambda_{111}\,, \lambda_{113})\to (\lambda_{hhh}^{\rm SM}\,, 0)$.
For these two limiting scenarios, further simplifications can be made for Eq.~\eqref{eq:dsigmahh_CPV}, which are
\beqs
\beqn
{\rm resonances}: \frac{d\hat \sigma }{d \hat t} &\approx& \frac{ G_F^2 \alpha_s^2 }{512\, (2\pi)^3}  \Big[ \Big| \sum_{h_i}  C_\triangle^{h_i} F_\triangle^h  \Big|^2 + \Big| \sum_{h_i} \tilde C_\triangle^{h_i} F_\triangle^A  \Big|^2  \Big]\,,\label{eq:dsigmahh_resonance}\\
{\rm non-resonances}: \frac{d\hat \sigma}{d \hat t}&\approx& \frac{G_F^2 \alpha_s^2 }{512\, (2\pi)^3} \Big[  \Big|   C_\triangle^{h_1}    F_\triangle^{h}  + C_\Box^{hh} F_\Box^{hh} + C_\Box^{AA} F_\Box^{AA}  \Big|^2 \non
&+&  \Big|  \tilde C_\triangle^{h_1}  F_\triangle^A  + C_\Box^{hA} F_\Box^{hA} \Big|^2  \non
&+& \Big| C_\Box^{hh} G_\Box^{hh} + C_\Box^{AA} G_\Box^{AA}  \Big|^2  + \Big|  C_\Box^{hA} G_\Box^{hA}  \Big|^2   \Big]\,.\label{eq:dsigmahh_nonresonance}
\eeqn
\eeqs
%

\begin{figure}
\centering
\includegraphics[width=6.5cm,height=4.5cm]{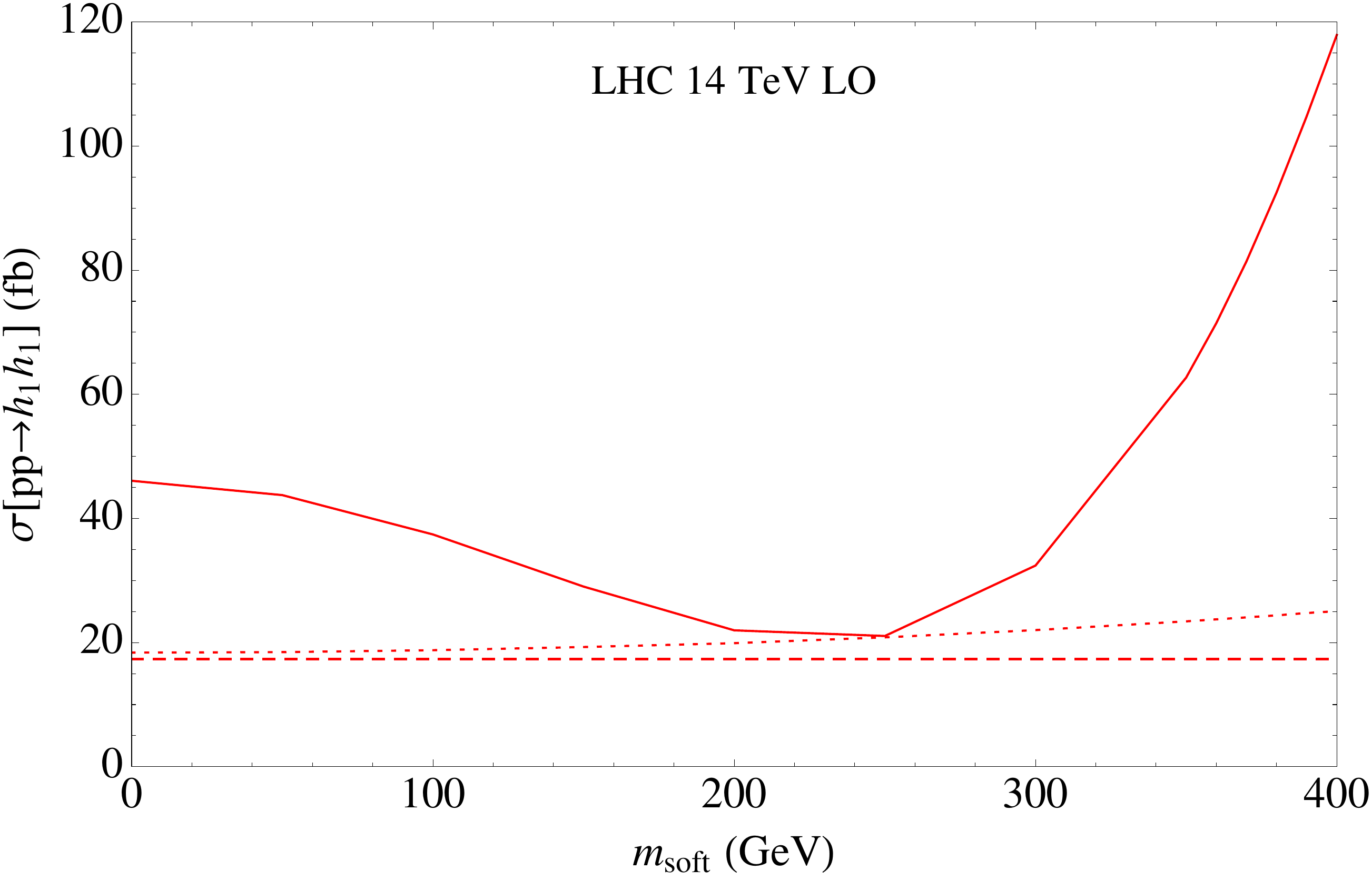}
\includegraphics[width=6.5cm,height=4.5cm]{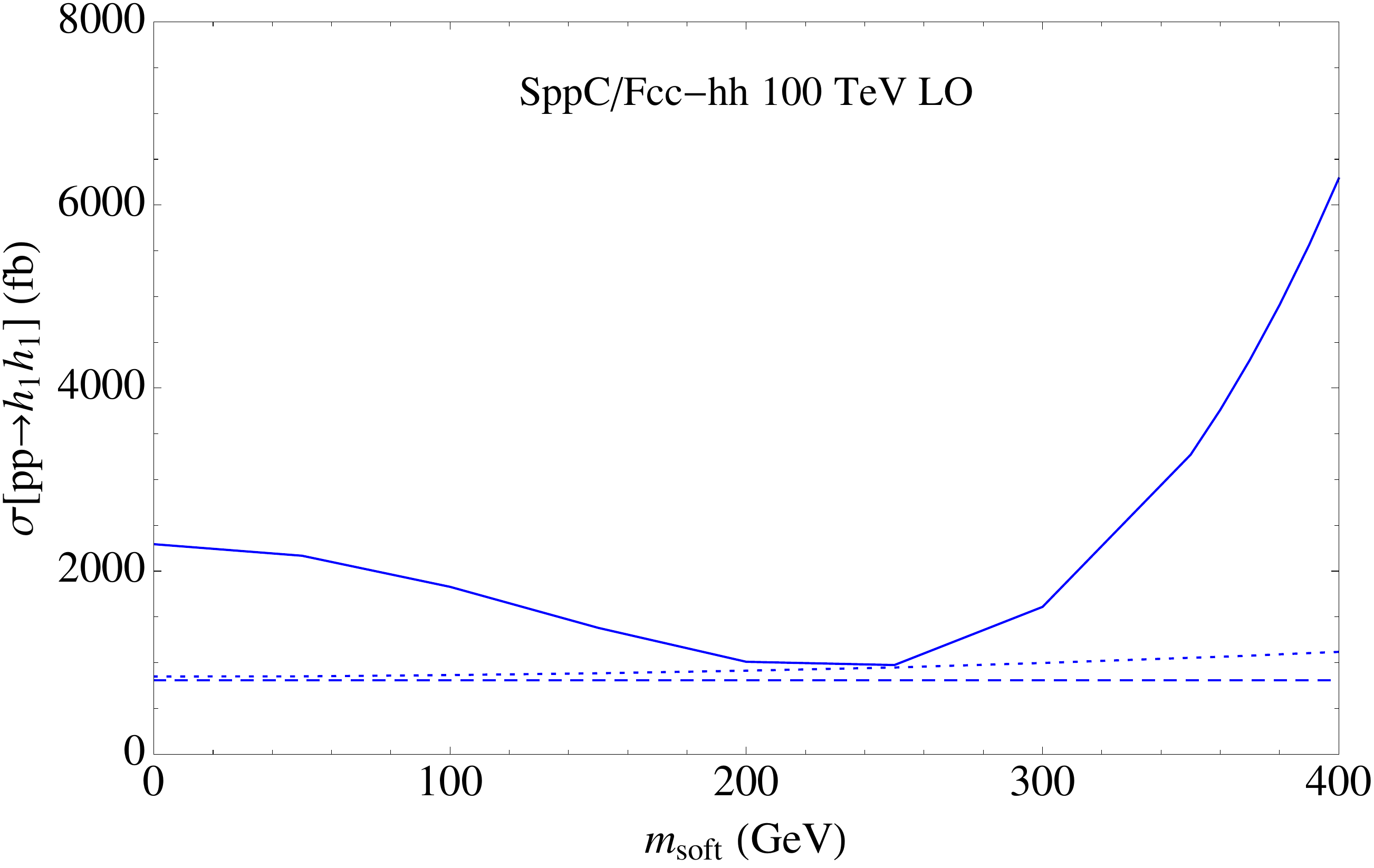}
\caption{\label{fig:sigmas} 
The cross sections of $\sigma[pp\to h_1 h_1]$ at the LHC $14\,\TeV$ (left) and SppC $100\,\TeV$ (right) versus the varying $m_{\rm soft}$ for the $M_2=M_3=600\,\GeV$ case in the CPV 2HDM-II, with fixed inputs of $|\alpha_b|=0.1$. 
}
\end{figure}

In Fig.~\ref{fig:sigmas}, we display the LO cross sections of $\sigma[pp\to h_1 h_1]$ at the LHC and the SppC/Fcc-hh for the $M_2=M_3=600\,\GeV$ case.
The solid curves represent the full results by combining every term in Eq.~\eqref{eq:dsigmahh_CPV}.
We also show the hypothetical cross sections of Eq.~\eqref{eq:dsigmahh_nonresonance}, where we turn off the Higgs cubic self coupling of $\lambda_{113}$ while modify $\lambda_{111}$ according to Eq.~\eqref{eq:lambda111}.
Thus, it is evident that the total cross sections approach to the SM-like Higgs pair productions with the modified cubic self couplings.
On the other hand, the LO cross sections at the LHC (SppC) can be as large as $\sim\mO(100)\,\fb$ ($\sim \mO(6)\,\pb$) when the soft mass approaches to the stability boundary for this case.

\begin{figure}[htb]
\centering
\includegraphics[width=6.5cm,height=4.5cm]{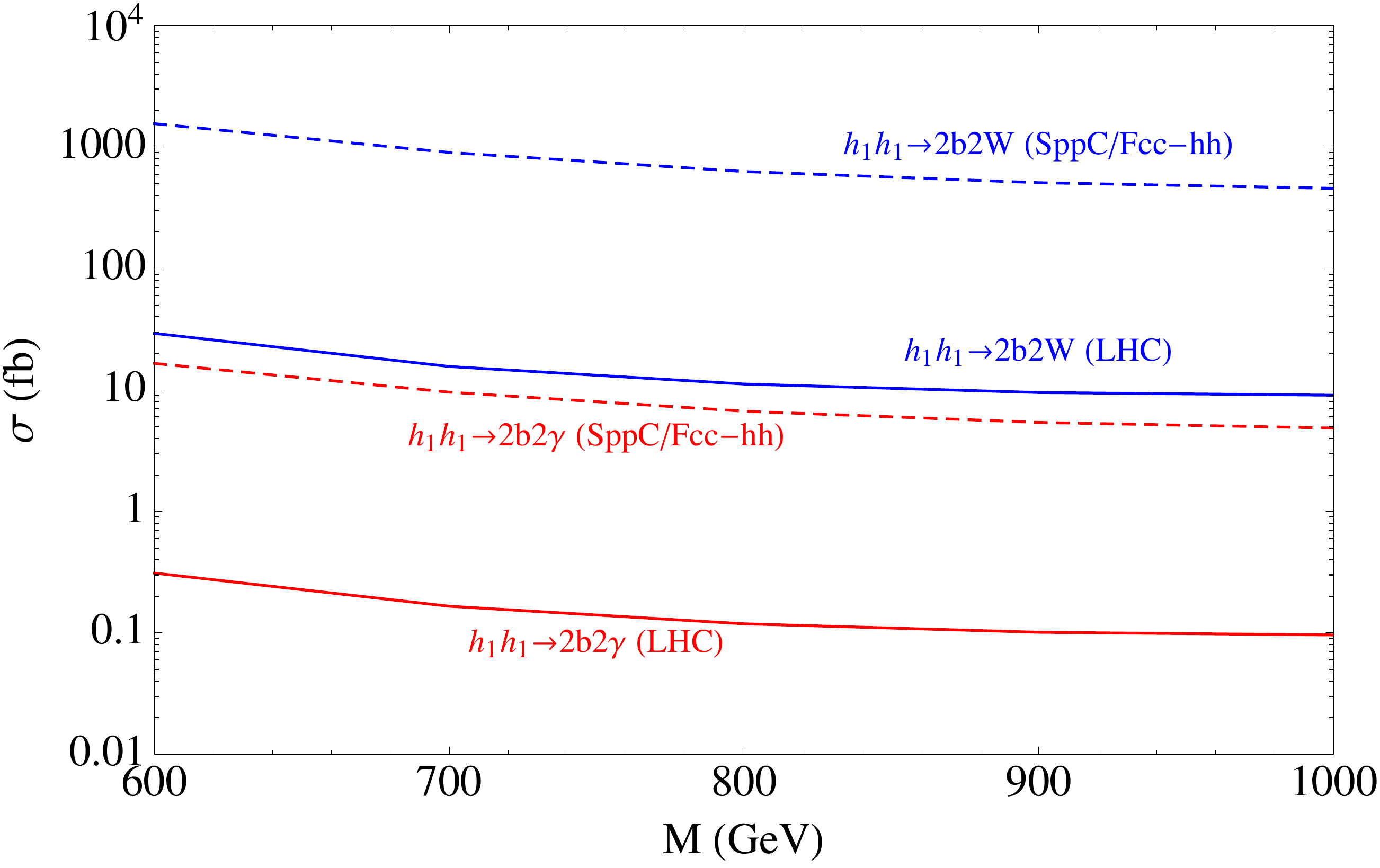}
\includegraphics[width=6.5cm,height=4.5cm]{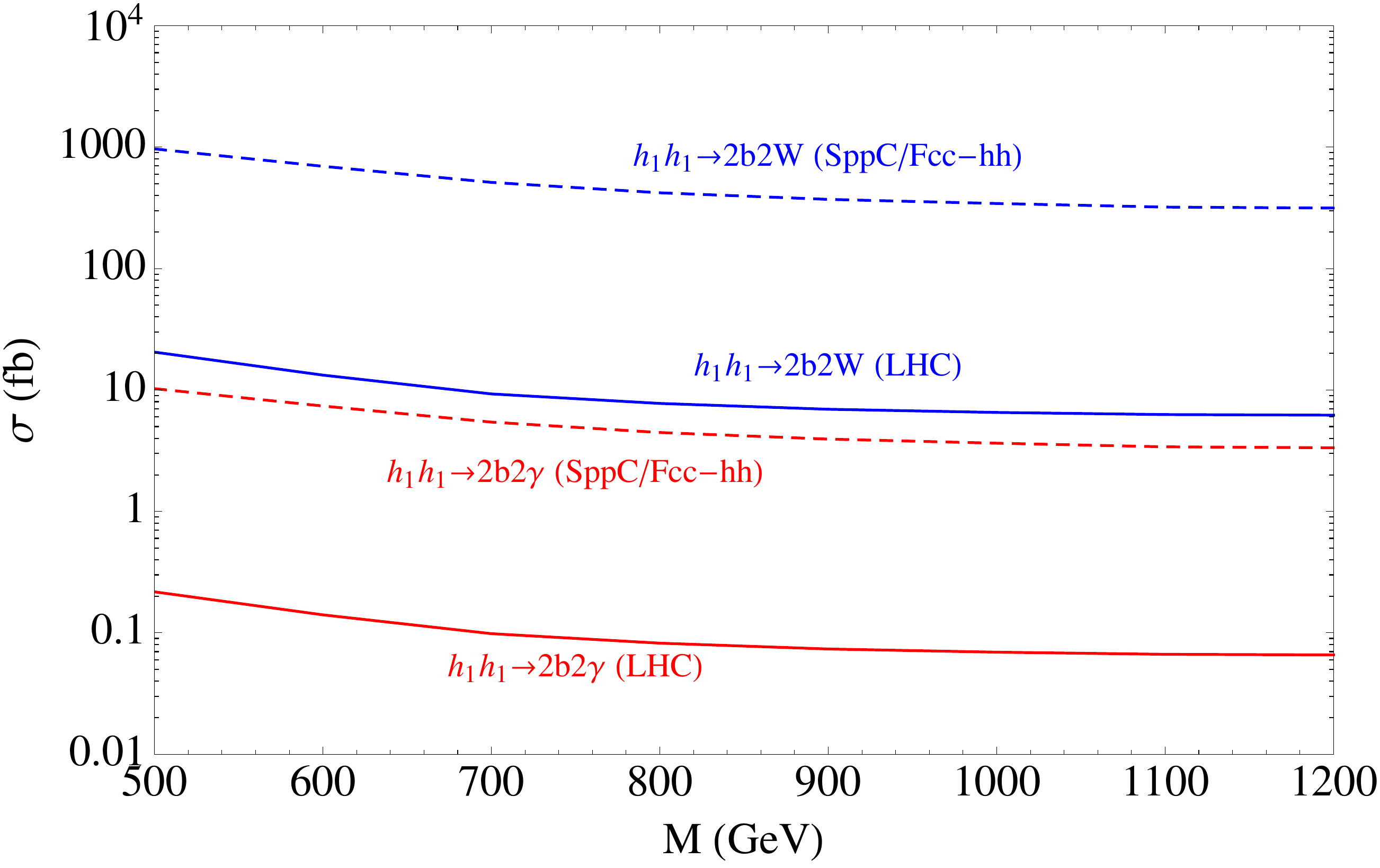}
\caption{\label{fig:ggh1h1}
The total cross sections of $\sigma[pp\to h_1 h_1]$ via the ggF at the LHC 14 TeV (solid curves) and the SppC 100 TeV (dashed curves).
Left: the cross sections for the benchmark models with the $\alpha_b=0.1$ input, right: the cross sections for the benchmark models with the $\alpha_b=0.05$ input.
}
\end{figure}

Furthermore, we evaluate the LO cross sections for the benchmark models listed in Table.~\ref{tab:benchmark}.
The typical cross sections subject all constraints in the previous context are $\mO(10)-\mO(100)\,\fb$ at the LHC or $\mO(1)\,\pb$ at the SppC for the allowed mass ranges.
The corresponding results are displayed in Fig.~\ref{fig:ggh1h1}, for benchmark models with $|\alpha_b|=0.1$ and $|\alpha_b|=0.05$, respectively.
We display the cross sections with the $h_1 h_1\to b \bar b + \gamma\gamma$ and $h_1 h_1 \to b \bar b + W^+ W^-$ final states.
From the experimental side, the $b \bar b + \gamma \gamma$ final states are the leading one to look for the Higgs pair productions at the hadron colliders, in that the relevant SM background is under control.
The LO cross sections for the $b \bar b + \gamma\gamma$ of our benchmark models are $\sim\mO(0.1)\,\fb$ at the LHC, and they increase to $\mO(10)\,\fb$ at the SppC.
In addition, one may also consider the $b \bar b + W_h W_\ell$ final states with the aid of the jet substructure technique~\cite{Papaefstathiou:2012qe}.
The LO cross sections for the $b \bar b + WW$ of our benchmark models are $\sim\mO(10)\,\fb$ at the LHC, and they increase to $\mO(1)\,\pb$ at the SppC.

\subsection{Other channels}

\begin{figure}[htb]
\centering
\includegraphics[width=6.5cm,height=4.5cm]{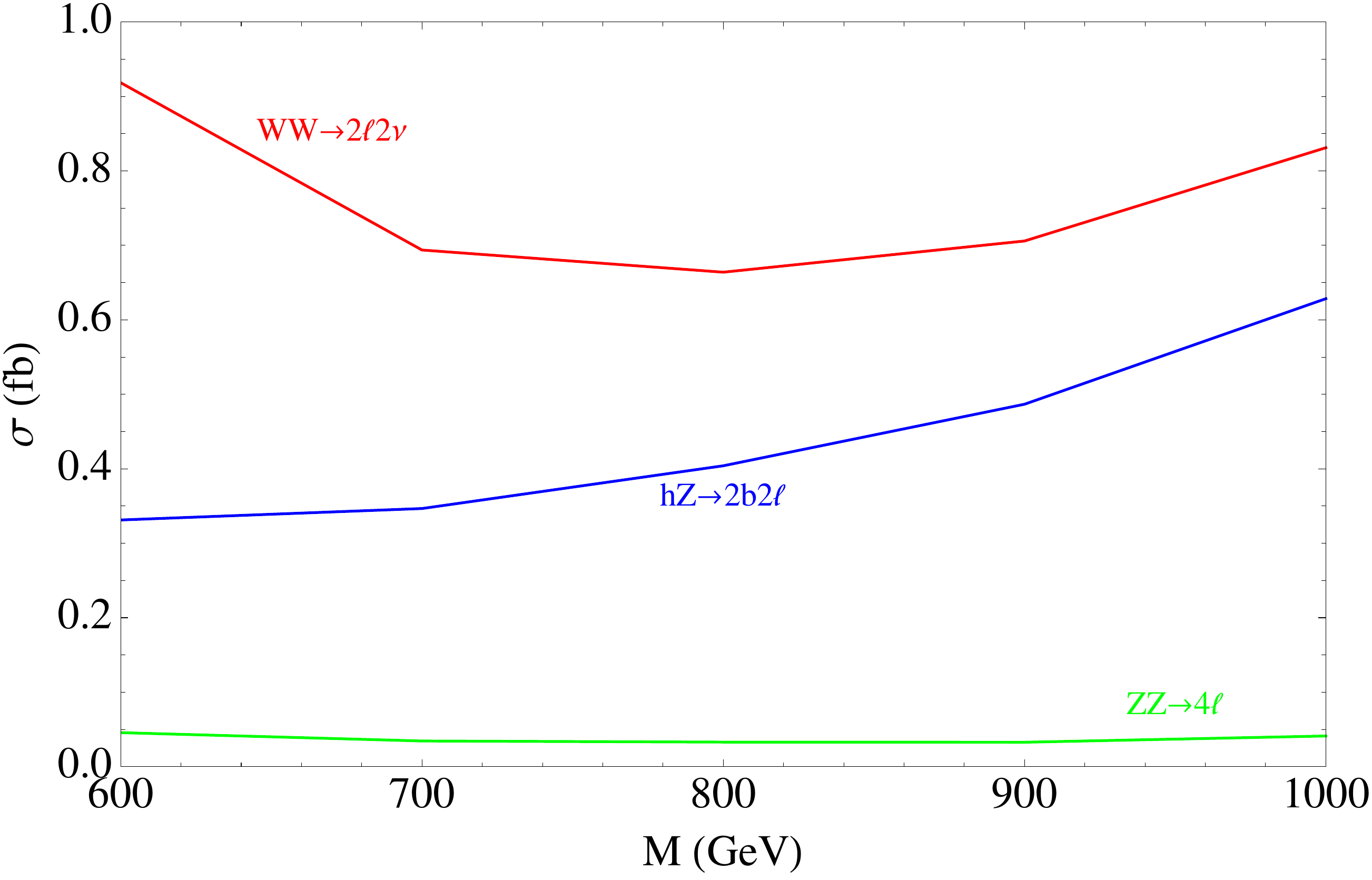}
\includegraphics[width=6.5cm,height=4.5cm]{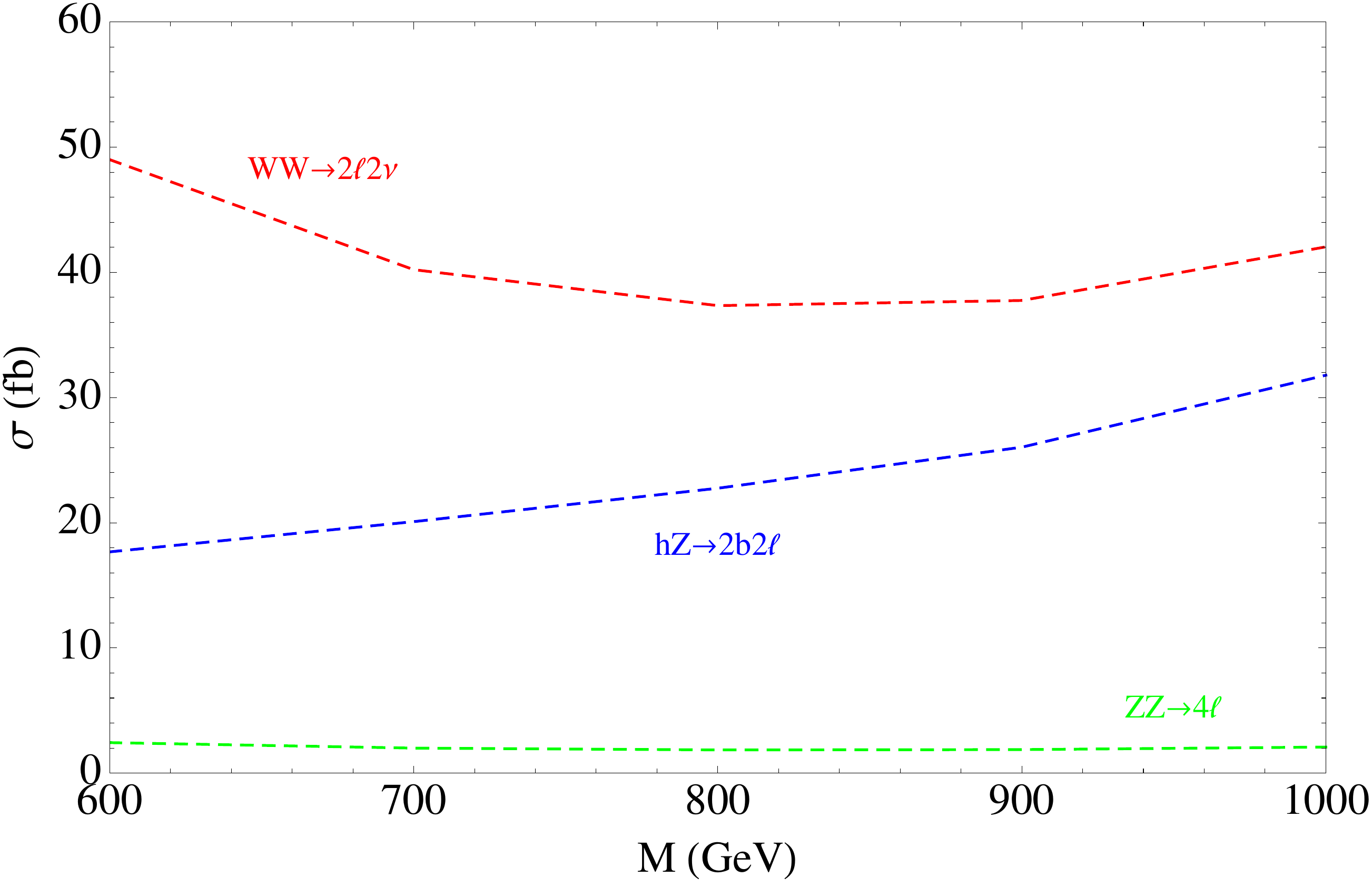}
\includegraphics[width=6.5cm,height=4.5cm]{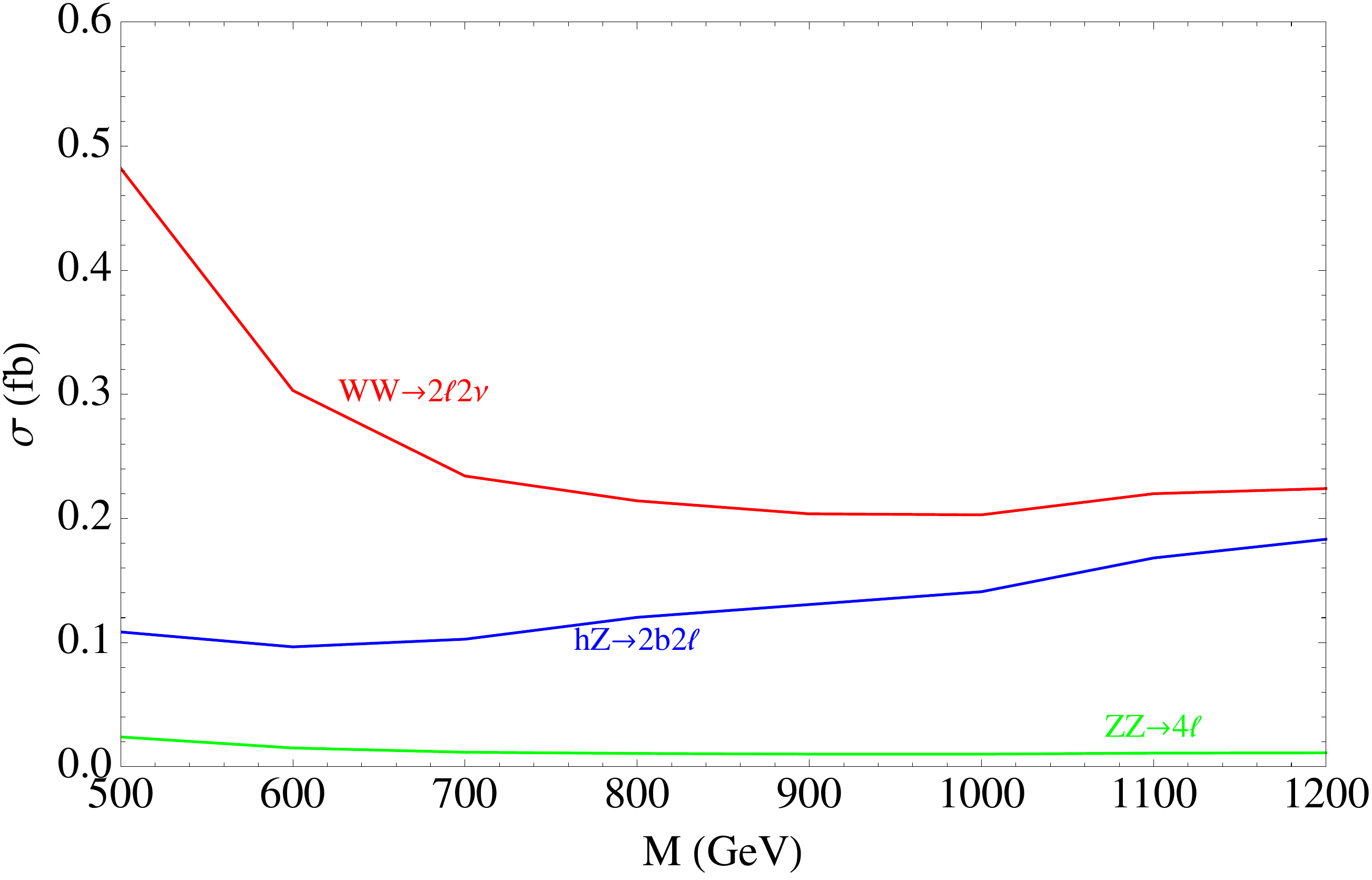}
\includegraphics[width=6.5cm,height=4.5cm]{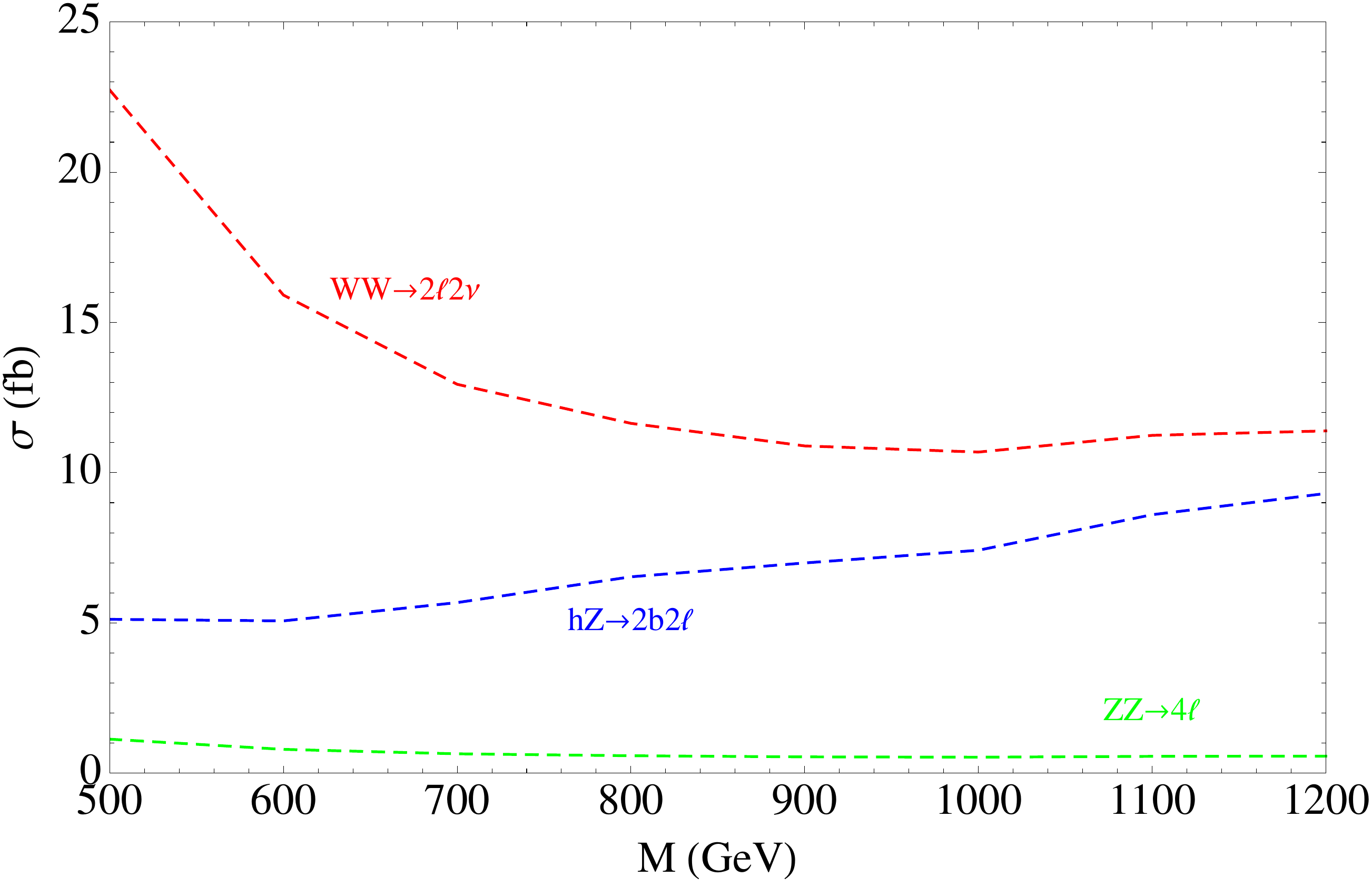}
\caption{\label{fig:WWZZhZ} 
The cross sections of the other search modes of the heavy Higgs bosons at the LHC 14 TeV (left panels) and the SppC 100 TeV (right panels).
}
\end{figure}

Besides the Higgs pair productions, we also have the other search modes for the heavy Higgs bosons of $h_2/h_3$, such as di-bosons and Higgs plus $Z$.
In Fig.~\ref{fig:WWZZhZ}, we display the cross sections of the other search modes of the heavy Higgs bosons, including $pp\to h_i\to (W^+ W^- \to 2\ell 2\nu \,,ZZ\to 4\ell \,,hZ\to b \bar b +\ell^+ \ell^-)$.
The current LHC searches for the heavy Higgs bosons via these channels can be found in Refs.~\cite{Khachatryan:2015cwa,Aad:2015kna,Khachatryan:2015tha,Aad:2015wra,Khachatryan:2015lba,Chatrchyan:2013yoa,TheATLAScollaboration:2013zha}.
The cross sections for these benchmark models are typically $\sim\mO(0.01)-\mO(0.1)\,\fb$ at the LHC, and enhanced to $\mO(1)-\mO(10)\,\fb$ at the SppC.
Analogous to the Higgs pair production process at the resonance region, the decay branching ratios of ${\rm Br}[h_i\to WW/ZZ/hZ] \propto \alpha_b^2$.
Therefore, the improvements of the precise measurements of the future eEDM experiments can also suppress the expected cross sections for these final states.


\section{Conclusion}
\label{section:conclusion}

The extended Higgs sector is a general setup with rich physical ingredients to address the issues that are beyond the SM. 
Particularly, the spontaneous CPV can be achieved with the general 2HDM setup.
In this work, we study the Higgs pair productions in the framework of the CPV 2HDM, with the focus on the leading production channel of the ggF.
The set of constraints to the CPV Higgs sector are taken into account, including the SM-like Higgs signal fit, the eEDM constraint, the perturbative unitarity and stability constraints, and the current LHC searches for the heavy Higgs bosons.
Together with the simplification to the model, we focus on the CPV 2HDM-II, where a relatively large size of CPV mixing is possible at $t_\beta\sim 1$.

The Higgs cubic self couplings play the most crucial role for the Higgs pair production.
For our case, two relevant cubic self couplings are $\lambda_{111}$ and $\lambda_{113}$, which are controlled by the soft mass term $m_{\rm soft}$ and the CPV mixing angle of $\alpha_b$.
The precise measurement of the SM-like Higgs cubic coupling of $\lambda_{111}$ can be achieved via the $e^+ e^- \to h_1 h_1 Z$ and $e^+ e^- \to h_1 h_1 \nu_e \bar \nu_e$ processes at the future high-energy $e^+ e^-$ colliders.
The benchmark models in our discussions typically predict totally cross sections of $\sigma[e^+ e^- \to h_1 h_1 Z]$ smaller than the SM predictions.
The largest deviations of the SM-like Higgs cubic couplings $\lambda_{111}$ are likely to be probed at the future TLEP $500\,\GeV$ and ILC $1\,\TeV$ runs.
At the future high-energy $pp$ collider runs, the Higgs pair productions are very likely to be controlled by the heavy resonance contributions. 
In the allowed mass range of the heavy Higgs bosons, we find the total production cross sections to be $\sigma[pp\to h_1 h_1]\sim \mO(10)- \mO(100)\,\fb$ at the LHC 14 TeV runs.
They can be as large as $\sim \mO(10^3)\,\fb$ at the future SppC 100 TeV runs.
Other search modes of di-bosons and Higgs plus $Z$ that are currently probed at the LHC $7\oplus 8\,\TeV$ experiments are also estimated at the future LHC $14\,\TeV$ and the SppC $100\,\TeV$ experiments.
The discovery of all these channels will manifest the structure of the Higgs sector.
Therefore, it will be very helpful to further study the higher-order QCD corrections as well as the collider search capabilities for such heavy resonance contributions to the Higgs pairs.


\section*{ACKNOWLEDGMENTS}

We would like to thank Jian Wang for his early collaboration in this work, and Yang Bai, Chien-Yi Chen, Jordy de Vries, Zuowei Liu, Lilin Yang and Yue Zhang for very useful discussions and communication. 
This work is partially supported by the National Science Foundation of China (under Grant No. 11575176, 11605016), the Fundamental Research Funds for the Central Universities (under Grant No. WK2030040069).
We would like to thank the Kavli Institute for Theoretical Physics China at the Chinese Academy of Sciences and Nanjing University for their hospitalities when part of this work was prepared.



\end{document}